\documentclass[12pt]{article}
\usepackage{jheppub}
\pdfoutput=1
\usepackage{epstopdf}

\usepackage{mathrsfs}
\usepackage{psfrag}
\usepackage{color}
\usepackage{caption}
\usepackage{subfig}
\usepackage[normalem]{ulem}
\usepackage{hyperref}
\usepackage[inline]{enumitem}
\usepackage{slashed}
\usepackage{feynmp-auto}
\usepackage{simplewick}
\usepackage{cancel}
\usepackage{multirow}
\usepackage{cleveref}
\usepackage{color}
\usepackage{shuffle}
\usepackage{environ}
\usepackage{tikz}
\usepackage{mathtools}
\usepackage{amssymb}
\usepackage{amsthm}
\usepackage{xspace}

\usetikzlibrary{shapes.geometric,arrows,arrows.meta,decorations.pathmorphing,decorations.markings,decorations.pathreplacing,patterns}

\makeatletter
\newsavebox{\measure@tikzpicture}
\NewEnviron{scaledtikzpicture}[1]{%
  \def\tikz@width{#1}%
  \def\tikzscale{1}\begin{lrbox}{\measure@tikzpicture}%
  \BODY
  \end{lrbox}%
  \pgfmathparse{#1/\wd\measure@tikzpicture}%
  \edef\tikzscale{\pgfmathresult}%
  \BODY
}
\makeatother

\crefname{figure}{figure}{figures}
\crefname{alg}{algorithm}{algorithms}

\newtheoremstyle{algo}{3pt}{3pt}{\normalfont}{}{\bfseries}{:}{ }{}
\theoremstyle{algo}
\newtheorem{alg}{Algorithm}

\DeclareMathOperator{\Pf}{Pf}
\DeclareMathOperator{\PT}{PT}
\DeclareMathOperator{\IT}{IT}
\DeclareMathOperator{\KK}{KK}

\newcommand{\Pfp}{\ensuremath{\text{Pf}\,'}}

\newcommand{\SEq}{\ensuremath{\overset{\text{SE}}{=}}}
\newcommand{\vf}{\ensuremath{\mathcal{V}}}

\newcommand{\lang}[1]{\ensuremath{\big<#1 \big|}}
\newcommand{\lsqr}[1]{\ensuremath{\big[#1 \big|}}
\newcommand{\rang}[1]{\ensuremath{\big|#1 \big>}}
\newcommand{\rsqr}[1]{\ensuremath{\big|#1 \big]}}

\newcommand{\SLC}{\ensuremath{\text{SL}(2,\mathbb{C})} }
\newcommand{\MMA}{{\sc Mathematica }}

\newcommand{\BF}[1]{\ensuremath{W_{\text{#1}}}}
\newcommand{\Part}{\textbf{Part}_2}
\newcommand{\Cdot}{\ensuremath{\!\cdot\!}}
\newcommand{\fs}[1]{\ensuremath{\slashed{f}_{\!#1}}}
\newcommand{\pol}{\ensuremath{\epsilon}}

\newcommand{\ord}{\ensuremath{\mathscr{O}}}

\newcommand{\crossing}{crossing symmetry\xspace}
\newcommand{\crossable}{crossing symmetric\xspace}
\newcommand{\Crossing}{Crossing symmetry\xspace}

\newcommand*{\halfway}{0.5*\pgfdecoratedpathlength+4.2pt}

\tikzset{dir/.style={decoration={markings, mark=at position \halfway with {\arrow{Latex}}},postaction={decorate}}}

\newcommand{\setarrowscale}[1]{
\tikzset{myarr/.style={decoration={markings,mark=at position \halfway with %
      {\arrow[scale=#1,>=Latex]{>}}},postaction={decorate}}}
}

\newcommand{\partsum}[2][(G,B)]{\ensuremath{\sum_{\substack{#1\in\Part(#2)}}}}

\definecolor{functioncolor}{rgb}{0,0,0}
\newcommand{\defn}[3]{~\\[-35pt]\begin{itemize}\item[]\indent\hspace{-21pt}$\bullet$\hspace{-.75pt} {\tt {\color{functioncolor}#1}\![}#2{\tt\,]\!:}#3\end{itemize}\vspace{-10pt}}

\newcommand{\defnn}[4]{~\\[-35pt]\begin{itemize}\item[]\indent\hspace{-21pt}$\bullet$\hspace{-.75pt} {\tt {\color{functioncolor}#1}\![}#2{\tt\,]\!\![}#3{\tt\,]\!:}#4\end{itemize}\vspace{-10pt}}

\definecolor{varcolor}{rgb}{0.1,0.55,0.25}
\newcommand{\vardef}[1]{{\color{varcolor} \texttt{#1}\_}}
\newcommand{\vardefH}[2]{{\color{varcolor} \texttt{#1}\hspace{-.5pt}\_\hspace{.5pt}\texttt{#2}}}
\newcommand{\vardefHH}[2]{{\color{varcolor} \texttt{#1}\hspace{-.5pt}\_\_\hspace{.5pt}\texttt{#2}}}
\newcommand{\varN}[1]{\textbf{({\color{varcolor} \texttt{#1}})}}

\title{Efficient Calculation of Crossing Symmetric BCJ Tree Numerators}

\author{Alex Edison}
\author{and Fei Teng}
\affiliation{Department of Physics and Astronomy, Uppsala University, SE-75108 Uppsala, Sweden}
\emailAdd{alexander.edison@physics.uu.se}
\emailAdd{fei.teng@physics.uu.se}

\preprint{UUITP--12/20}

\abstract{ In this paper, we develop an improved method for directly
  calculating double-copy-compatible tree numerators in
  (super-)Yang-Mills and Yang-Mills-scalar theories.  Our
  new scheme gets rid of any explicit dependence on \emph{reference
    orderings}, restoring a form of \crossing to the numerators.  This
  in turn improves the computational efficiency of the algorithm,
  allowing us to go well beyond the number of external particles
  accessible with the reference order based methods.  Motivated by a
  parallel study of one-loop BCJ numerators from forward limits, we
  explore the generalization to include a pair of fermions.  To
  improve the accessibility of the new algorithm, we provide a \MMA
  package that implements the numerator construction.  The structure
  of the computation also provides for a straightforward introduction
  of minimally-coupled massive particles potentially useful for future
  computations in both classical and quantum gravity.
}

\begin{document}
\maketitle

\section{Introduction}
The efficient calculation of scattering amplitudes at high precision
relies on the discovery of novel methods and structures, many of which
are shared between disparate theories.  While these structures are
often highly obscured by traditional Lagrangian/Feynman diagram
approaches, they tend to lead to new formulations of amplitudes in
quantum field theories.

One powerful structure in tree amplitudes is the
Bern-Carrasco-Johansson (BCJ) color-kinematics duality
\cite{Bern:2008qj,Bern:2010ue,Bern:2019prr}, which allows a large class of
theories to be constructed as the double copy of simpler theories.
This duality has been used to push the envelop on accessible
calculations in various theories of quantum gravity
\cite{Johansson:2017bfl,Bern:2018jmv,Bern:2014sna,Bern:2013uka} by
``squaring'' (super-)Yang-Mills (sYM) theory.  Additionally,
BCJ-compatible trees have been used in the blossoming application of
scattering amplitudes techniques to black hole physics in classical
gravity~\cite{Luna:2016due,Goldberger:2016iau,Luna:2017dtq,Shen:2018ebu,Plefka:2018dpa,Cheung:2018wkq,Kosower:2018adc,Guevara:2018wpp,Bern:2019nnu,Plefka:2019hmz,
  Maybee:2019jus,Guevara:2019fsj,Johansson:2019dnu,Bern:2019crd,Bautista:2019evw,Plefka:2019wyg}.
Unfortunately, most textbook methods for calculating tree amplitudes
do not directly generate color-kinematics dual representations.

The Cachazo-He-Yuan (CHY) formalism
\cite{Cachazo:2013gna,Cachazo:2013hca,Cachazo:2013iea} is a powerful
method for studying the properties of tree-level scattering amplitudes
in many theories, in part because it manifests double-copy properties.
The formalism explicitly constructs tree amplitudes from the product
of two half-integrands. There is now a well-developed program for
exploring the types of objects needed in each half-integrand to
generate theories of
interest~\cite{Cachazo:2014nsa,Cachazo:2014xea,He:2016iqi,Azevedo:2017lkz,Cachazo:2018hqa,He:2018pol,Geyer:2018xgb,He:2019drm},
which can be derived from world-sheet models based on ambitwistor
strings~\cite{Mason:2013sva,Berkovits:2013xba,Adamo:2013tsa,Adamo:2015hoa,Casali:2015vta}
and their deformation~\cite{Azevedo:2017yjy,Azevedo:2019zbn}.  The
half-integrands are known to be expressible completely in Del
Duca-Dixon-Maltoni half-ladder basis \cite{DelDuca:1999rs} on the
support of scattering equations~\cite{Cachazo:2013iea},\footnote{This expansion also holds for worldsheet correlators in generic string theories~\cite{Mafra:2011kj,Broedel:2013tta,Huang:2016tag,Mizera:2017cqs,Mizera:2017rqa,Azevedo:2018dgo}.}
\begin{align}\label{eq:ddmExp}
  \mathcal{I}_n\overset{\text{SE}}{=}\sum_{\beta\in S_{n-2}}\PT(1,\beta,n) N(1,\beta,n)\,,\,\text{where } \PT(\rho)\equiv\frac{1}{\sigma_{\rho_1\rho_2}\sigma_{\rho_2\rho_3}\cdots\sigma_{\rho_{n-1}\rho_n}\sigma_{\rho_n\rho_1}}\,.
\end{align}
In this basis, the kinematic coefficients $N(1,\beta,n)$ of the
worldsheet \emph{Parke-Taylor factors} $\PT(1,\beta,n)$ can be seen as
master numerators, since all non-ladder kinematic numerators can be
built using BCJ numerator relations on the
half-ladders~\cite{Bern:2010yg}. Due to this fact, the half-ladder
numerators fully specify tree-level amplitudes.  This can be
explicitly seen when considering sYM through the use of the
\emph{doubly-color-ordered biadjoint scalar amplitudes}
$m(\alpha|\beta)$
\cite{Vaman:2010ez,Du:2011js,BjerrumBohr:2012mg,Cachazo:2013iea},
which combine the half-ladder numerators $N$ and the DDM color
structures $c$ \cite{DelDuca:1999rs} with the appropriate signs and
propagators to generate amplitudes via
\begin{equation}
  \mathcal{A}_n^{\text{tree}} = \mathscr{N} \sum_{\alpha,\beta \in S_{n-2}} c(1,\alpha,n)m(1,\alpha,n|1, \beta,n) N(1,\beta,n) \,.
  \label{eq:ym-amp}
\end{equation}
These numerators can be directly double-copied with appropriate
partners $\tilde{N}$ to yield various gravitational theories,
\begin{equation}
  \mathcal{M}_n^{\text{tree}} = \mathscr{N}_{\text{GR}}\sum_{\alpha,\beta \in S_{n-2}} \tilde{N}(1,\alpha,n)m(1,\alpha,n|1, \beta,n) N(1,\beta,n) \,. \label{eq:grav-amp}
\end{equation}
As amplitudes for bosonic theories, both $\mathcal{A}$ and
$\mathcal{M}$ are \emph{invariant} under particle exchange.  The
permutation invariance of $2,\dots,n{-}1$ is manifest if
\begin{align}
  N(1,\beta,n)\equiv N(1,2,\ldots,n)\Big|_{2\rightarrow\beta_2,3\rightarrow\beta_3,\ldots,n{-}1\rightarrow\beta_{n-1}}\,,
  \label{eq:crossing}
\end{align}
which we will take as our definition of \emph{crossing symmetry}
acting on numerators.\footnote{The crossing symmetry in $1$ and $n$
  are due in part to the BCJ relations.  Since the $N$ are for
  half-ladders, one might expect that exchanging either
  $1\leftrightarrow\beta_2$ or $n\leftrightarrow\beta_{n-1}$ in
  $N(1,\beta,n)$ would yield a minus sign.  However, the interplay
  between color and kinematic jacobis allows this property of the
  half-ladders to be broken or restored as needed. Of course, imposing these additional \crossing will further eliminate the gauge freedom in these numerators. On the other hand, when $1$
  and $n$ are a different species from the rest, \crossing involving
  them is necessarily absent.}  The resemblance of \cref{eq:grav-amp}
to the Kawai-Lewellen-Tye (KLT) double-copy
relation~\cite{Kawai:1985xq} provides a direct construction of
$N(1,\beta,n)$ using color-ordered amplitudes and KLT momentum
kernel~\cite{KiermaierTalk,BjerrumBohr:2010hn}. However, the result is
non-local as the numerator is a rational function of Mandelstam
variables, rather than polynomial.

On the other hand, CHY formalism provides a way to compute
\emph{local} numerators in the DDM basis through
\cref{eq:ddmExp}. Although the possibility was recognized shortly
after the discovery of CHY~\cite{Cachazo:2013iea}, the explicit construction was highly
nontrivial and came much later.\footnote{See Ref.~\cite{Bjerrum-Bohr:2016axv} for a generic construction of \emph{non-local} BCJ numerators using CHY integration rules.} The computation for up to three gluons and two traces in Yang-Mills-scalar (YMS) theory~\cite{Chiodaroli:2014xia} was done
in~\cite{Stieberger:2016lng,Nandan:2016pya,delaCruz:2016gnm,Schlotterer:2016cxa}. Shortly
after, a recursive expansion for the single-trace sector of
YMS  was proposed based on gauge
invariance~\cite{Fu:2017uzt} and a direct ansatz construction~\cite{Chiodaroli:2017ngp}, which was soon proved and refined by
expanding the CHY integrand recursively~\cite{Teng:2017tbo}. Follow-up
works further extended the algorithm to pure
Yang-Mills~\cite{Du:2017kpo} and multi-trace
sector~\cite{Du:2017gnh}. A crucial feature common to all these
constructions is that the DDM basis numerators are given by assigning
kinematic factors to \emph{spanning trees} on $n$ vertices ---
connected graphs on $n$ vertices with no cycles. 

While the spanning tree construction is generic, the need for a
\emph{reference ordering} (RO) causes the numerators to be a function
of more than just the particle ordering, and thus not manifest
\crossing,
\begin{align}
  N_{\text{RO}}(1,\beta,n)\neq N_{\text{RO}}(1,2,\ldots,n)\Big|_{2\rightarrow\beta_2,3\rightarrow\beta_3,\ldots,n{-}1\rightarrow\beta_{n-1}}\,.
  \label{eq:not-cross}
\end{align}
Although this is not in line with \cref{eq:crossing}, the full
amplitude remains permutation invariant due to the nontrivial kernel
of $m(1,\alpha,n|1,\beta,n)$, {\it i.e.} the BCJ amplitude relations.
\Crossing can be recovered by \emph{averaging over all the ROs}
(RO-average). However, this prescription introduces a factorial
computational complexity, which is highly undesired. We note that the
mathematical origin of this RO dependence roots in the Laplace
expansion of determinants and Pfaffians: when one performs the
expansion along a chosen row, the manifest exchange symmetry between
this row and the rest is broken. Different choices lead to different
patterns of breaking the symmetry. Alternatively, the prescriptions
presented in~\cite{Fu:2017uzt,Du:2017kpo,Du:2017gnh,Teng:2017tbo} can
be derived from conventional string
theory~\cite{Schlotterer:2016cxa,He:2018pol,He:2019drm} and differential operators~\cite{Feng:2019tvb}. The stringy interpretation of the
RO is a priority list for breaking the subcycles formed by world-sheet
variables using integration-by-parts relations.

In section~\ref{sec:chy} and part of section~\ref{sec:expand},
  we will review the RO-based construction of sYM and multi-trace YMS
  numerators. While the technique summarized in \cref{al:ro} is mostly equivalent to those in
  Refs.~\cite{Fu:2017uzt,Du:2017kpo,Du:2017gnh,Teng:2017tbo,He:2019drm},
  we streamline the algorithm with new notations that are more
  suitable to describe the novel \crossable contruction discussed
  later. In addition, we propose a new baseline
  expansion~\eqref{eq:BLExpMT} for YMS that works for any choice of
  particle $1$ and $n$ (gluon, scalars in the same or different
  traces).\footnote{It is a natural generalization of
    Ref.~\cite{Du:2017gnh}, where $1$ and $n$ are fixed to be scalars.}

While the lack of \crossing does not pose a problem when studying
tree-level amplitudes, it can be challenging to deal with when using
trees as part of loop calculations.  The inability to relabel
numerators prevents systematic approaches using graph isomorphisms
that might generalize to loops. However, the CHY formalism has been
extended to handle one-loop directly.  This representation of loop
amplitudes can be derived from ambitwistor string models based on a
nodal Riemann
sphere~\cite{Adamo:2013tsa,Geyer:2015bja,Geyer:2015jch,Geyer:2016wjx},
which in turn allows direct calculation from an $(n{+}2)$-point
tree-level amplitude through a forward
limit~\cite{Cachazo:2015aol,He:2015yua}. Following this approach, one
can avoid the difficulty of dealing with contributions from different
spin structures at one-loop level. The resultant loop integrands
have the prominent feature that they contain linearized propagators
rather than the standard Feynman-type ones. Moreover, kinematic
numerators satisfying BCJ relations will automatically lead to such
numerators at one-loop, which can be directly fed into a double-copy
construction for gravity
theories~\cite{He:2016mzd,He:2017spx}. These method of obtaining
BCJ-compatible one-loop numerators have been explored by various works
in
literature~\cite{He:2016mzd,He:2017spx,Geyer:2017ela,Agerskov:2019ryp}.

Thus, high multiplicity loops in many theories are accessible from
high multiplicity trees via CHY and forward limits.  The first
technical difficulty to overcome towards this goal is the need to
efficiently produce tree-level BCJ numerators with very high
multiplicity. As the main goal of this paper, we develop an improved
algorithm to arrive at \crossable numerators \emph{without} performing
the explicit after-the-fact RO-average. In other words, the new method
gets rid of RO-dependence completely.  The procedure,
  presented in \cref{al:rec}, works by processing the structure of a
  spanning tree \emph{recursively} rather than sequentially over
  arbitrarily split paths.  The recursive structure allows kinematic
  weights to be assigned at each node which locally incorporate the
  appropriate tensor structures and normalizations such that a
  spanning tree processed in the new algorithm gives the same result
  as performing the RO-average after the old construction
  \cite{Fu:2017uzt,Du:2017kpo,Du:2017gnh,Teng:2017tbo,He:2019drm}. The
  bulk of this paper is dedicated to examining the combinatorics and
  kinematic factors needed at each stage of the recursion. In
  section~\ref{sec:rec-constr}, we derive the prescription for sYM
  amplitudes containing at most two fermions.\footnote{Numerators with
    more fermions can be obtained from the pure-spinor formalism of
    sYM~\cite{Mafra:2011kj,Mafra:2015vca}.} We then generalize it to
  multi-trace sector of YMS theory in section~\ref{sec:rec-con-mt}.
Our algorithm will automatically respect the \crossing among gluons
and scalar traces, all of which are not manifest when a RO is
required. We have checked that our algorithm can provide complete
ten-point results within minutes on a laptop. In a forthcoming
paper~\cite{Edison:2020uzf}, we will feed these tree-level numerators
into the forward-limit machinery to study one-loop numerators with
various amount of supersymmetry.

Additionally, since our formulation is dimensionally agnostic, it is
possible to embed the numerators in higher dimensions in such a way as
to introduce mass to specific particles via dimensional compactification.  In
particular, it is straightforward to perform this embedding such that
only particles $1$ and $n$ obtain a mass.  These types of diagrams are
related to those needed for the recent 3PM (third post-Minkowski)
calculation of Ref.~\cite{Bern:2019crd} as input to the unitarity cuts
used for constructing the two-loop classical amplitude.

Our paper is organized as follows. In \cref{sec:chy}, we review the
general structure of the CHY representation of integrands.  This
section specifically focuses on the initial expansion of $\Pfp$ into
pure $\Pf$ weighted by kinematics and Parke-Taylor factors.  We cover
pure gluon, two-fermion, and multi-trace YMS.  Section
\ref{sec:expand} begins by reviewing the spanning tree method for CHY
(half-)integrand construction.  It then presents our derivation of a
recursive restructuring of the spanning tree calculation which
implicitly performs an average over reference ordering, as well as
providing some examples of the new technique. Finally, in
\cref{sec:mass}, we briefly discuss introducing masses to our
numerators. Appendix \ref{sec:CHYtoAmp} provides our explicit
conventions for the CHY matrices and normalizations.  Appendix
\ref{sec:package} describes our \MMA package that implements the
spanning tree numerator construction.

\section{CHY Integrands and Baseline Expansion}
\label{sec:chy}
In the CHY representation, $n$-point tree-level color-ordered
amplitudes are expressed as an integral over the moduli space of
$n$-punctured Riemann sphere,
\begin{align}\label{eq:CHY}
A^{\text{tree}}_n(\rho)=\mathscr{N}\int d\mu_n^{\text{tree}}\PT(\rho)\,\mathcal{I}_n^{\text{tree}}\,,& &d\mu_n^{\text{tree}}\equiv\frac{d^n\sigma}{\text{Vol}[\text{SL}(2,\mathbb{C})]}\sideset{}{^{\,\prime}}\prod_{i=1}^{n}\delta(E_i)\,,
\end{align}
where the delta functions in the measure localize the integral to the
$(n-3)!$ solutions of the scattering
equations~\cite{Cachazo:2013gna,Cachazo:2013hca,Cachazo:2013iea},
\begin{align}
  E_i\equiv\sum_{\substack{j=1 \\ j\neq i}}^{n}\frac{k_i\Cdot k_j}{\sigma_{ij}}=0\,,& &\sigma_{ij}\equiv\sigma_i-\sigma_j\,.
\end{align}
The color ordering is encoded in the Parke-Taylor factor $\PT(\rho)$,
which is a weight-two function under the \SLC transformation
$\sigma_{i}\rightarrow\frac{a\sigma_i+b}{c\sigma_i+d}$ with
$ad-bc=1$. The (half-)integrand $\mathcal{I}_n^{\text{tree}}$ depends
on the kinematic data (momentum and polarization). It is constructed
as an \SLC weight-two function as well such that the full integration
form is \SLC invariant. The \SLC gauge redundancy, combined with the
$\frac{1}{\text{Vol}[\text{SL}(2,\mathbb{C})]}$, can be removed by
fixing the positions of three punctures and dropping three redundant
scattering equations. For example, we can choose
\begin{align}\label{eq:wsgf}
  d\mu_n^{\text{tree}}=(\sigma_{1,n-1}\sigma_{n-1,n}\sigma_{n1})^2\prod_{i=2}^{n-2}\Big[d\sigma_i\,\delta(E_i)\Big]\,,
\end{align}
where $(\sigma_1,\sigma_{n-1},\sigma_n)=(0,1,\infty)$. The
normalization $\mathscr{N}$ contains the coupling constants of the
theory, which we match against the explicit Lagrangian in \cref{sec:CHYtoAmp}.

\subsection{Super-Yang-Mills (sYM) theories}
\label{sec:sym}

We first focus on sYM theories. The CHY integrand for pure gluon
amplitudes is~\cite{Cachazo:2013hca}
\begin{align}\label{eq:gluon}
  \mathcal{I}_n^{\text{gluon}}=\Pfp(\Psi)\,.
\end{align}
The full definition of the \emph{reduced Pfaffian} $\Pfp(\Psi)$ is given in \cref{sec:CHYtoAmp}. One can rewrite $\Pfp(\Psi)$ as a \emph{baseline expansion} on
the support of the scattering
equations~\cite{Lam:2016tlk,Fu:2017uzt,Du:2017kpo},
\begin{align}\label{eq:BLExp}
  \mathcal{I}_n^{\text{gluon}}\,\SEq\partsum{2,3,\ldots,n-1}\Pf(\Psi_G)\sum_{\rho\in S_{|B|}}\PT(1,\rho,n)\BF{gluon}(1,\rho,n)\,,
\end{align}
where the first summation is over all the bi-partitions of the set
$\{2,3,\ldots,n-1\}$, and the second summation is over all the
permutations of the set $B$.\footnote{This sum is often written in the
  literature as over $\{2, \ldots, n-1\} = A \cup B$, but we wish to avoid
  abuse of notation since $A$ and $B$ are required to be
  disjoint.} The matrix $\Psi_G$ is a submatrix of $\Psi$ in which
the rows and columns are restricted to the set $G$ (see
\cref{sec:CHYtoAmp} for more details). We note that both $G$ and $B$
can be empty. In particular, when $G=\emptyset$,
$\Pf(\emptyset)=1$. We call $(1,\rho,n)$ a \emph{baseline} and the
\emph{baseline factor} $\BF{gluon}$ is given by
\begin{align}
  \BF{gluon}(1,\rho,n)=(-1)^{|\rho|}\pol_1\Cdot f_{\rho_1}\Cdot f_{\rho_2}\Cdot\ldots\Cdot f_{\rho_{|B|}}\Cdot\pol_n\,,
\end{align}
where
$f_{i}^{\mu\nu}\equiv k_i^{\mu}\pol_i^{\nu}-\pol_i^{\mu}k_i^{\nu}$ is
the linearized field strength for gluon $i$. Our nomenclature is
inspired by the \emph{spanning tree} expansion scheme for $\Pfp$,
which will be discussed in \cref{sec:expand}. For the derivation of \cref{eq:BLExp}, we refer the interested readers to the aforementioned references. From the ambitwistor
string point of view, singling out gluon $1$ and $n$ amounts to
assigning ghost picture $-1$ to the vertex operators in the RNS
formalism~\cite{Mason:2013sva}. The gauge invariance of these two
particles are not manifest, but can be recovered by using the
scattering equations.

Interestingly, the CHY integrand with exactly two external fermions
takes a similar baseline expansion form. If we fix the fermions to be
particle $1$ and $n$, the only difference from \cref{eq:BLExp} is in
the baseline factor~\cite{Edison:2020uzf},
\begin{align}\label{eq:2fExp}
\mathcal{I}_n^{\text{2f}}(1_{\text{f}},2,\ldots,n{-}1,n_{\text{f}})=\partsum{2,3,\ldots,n-1}\Pf(\Psi_G)\sum_{\rho\in S_{|B|}}\PT(1,\rho,n)\BF{2f}(1_{\text{f}},\rho,n_{\text{f}})\,.
\end{align}
There exist two forms of the baseline factor $\BF{2f}$, equivalent on
the support of the scattering equations and gauge transformations. In $D=10$, they are given as\footnote{The
  two expressions originate from the two ways to assign ghost picture
  to the vertex operators in the RNS formalism of ambitwistor string:
  \cref{eq:WF1} comes from assigning ghost picture $-1/2$ to both
  fermion vertex operators and ghost picture $-1$ to the gluon vertex
  operator $m$, while \cref{eq:WF2} comes from assigning ghost picture
  $-1/2$ and $-3/2$ to the fermion vertex operators and ghost picture
  zero to all the gluons.  We have found the two expressions useful
  for studying different aspects of forward limits, see
  Ref.~\cite{Edison:2020uzf} for details.}
\begin{subequations}
\begin{align}
\label{eq:WF1}
  \BF{2f}^{(1)}(1_{\text{f}},\rho,n_{\text{f}})&=\left\{\begin{array}{ccc}\displaystyle
\frac{(-1)^{|\rho|}}{2}(\chi_1\fs{\rho_1}\ldots\fs{\rho_{i-1}}\slashed{\pol}_{\ell}\fs{\rho_{i+1}}\ldots\fs{\rho_{|B|}}\chi_n) &\quad & \rho_i=\ell \\
0 &\quad & \ell\notin\rho 
\end{array}\right., \\
\label{eq:WF2}
  \BF{2f}^{(2)}(1_{\text{f}},\rho,n_{\text{f}})&=(-1)^{|\rho|}(\chi_1\fs{\rho_1}\fs{\rho_2}\ldots\fs{\rho_{|B|}}\xi_n)\,,
\end{align}
\end{subequations}
where in the first equation, $\ell$ is an arbitrary gluon and the
baseline factor vanishes if $\ell\notin B$. Eq.~(\ref{eq:WF2}) has the
benefit of being manifestly crossing symmetric and gauge invariant in
all of the gluons. In both expressions, the gamma-matrix convention is
$\{\gamma^{\mu},\gamma^{\nu}\}=2\eta^{\mu\nu}$, and
\begin{align}
  \slashed{\pol}_\ell\equiv\pol_\ell^{\mu}\gamma_{\mu}\,,& &\fs{i}\equiv\frac{1}{8}f_{i}^{\mu\nu}[\gamma_{\mu},\gamma_{\nu}]=\frac{1}{2}\slashed{k}_i\slashed{\pol}_i\,.
\label{eq:tr-f}
\end{align}
The Weyl-Majorana spinor $\chi_i$ is the solution to the equation of
motion $\slashed{k}_i\chi_i=0$, and the spinor $\xi_i$ is related to
$\chi_i$ through $\slashed{k}_i\xi_i=\chi_i$. The expressions for
$D<10$ can be obtained through a dimensional reduction. In particular,
for $D=4$ we can use
\begin{align}\label{eq:dimred}
\chi_i\rightarrow  \lang{i}^{\alpha,I}\oplus \lsqr{i}_{\dot{\alpha},I}\,,& &\xi_i\rightarrow \frac{\rang{q}_{\alpha,I}}{\langle iq\rangle}\oplus\frac{\rsqr{q}^{\dot{\alpha},I}}{[iq]}
\end{align}
where $\alpha,\dot{\alpha}=1,2$ are left- and right-handed Weyl spinor
indices, $I=1,2,3,4$ is the $\text{SU}(4)$ R-symmetry index and the $\oplus$
is a formal sum over the representation spaces.  In $D=6$, we can
similarly construct $\chi$ and $\xi$ using 6D spinor helicity
variables~\cite{Cheung:2009dc}
\begin{align}
  \chi_i \to \rang{i_{a}}^A \oplus \rsqr{i_{\dot a}}_A\,,  &&  \xi_i \to \frac{\rsqr{q_{\dot{b}}}_A  \big[ q^{\dot b} i_{a}\big>}{2q\Cdot k_i}  \oplus
                                                                 \frac{\rang{q^{b}}^A  \left< q_b i_{\dot{a}}\right]}{2q\Cdot k_i}
\label{eq:sixd-ferm}
\end{align}
where the spinors $\rang{i}^A$ and $\rsqr{i}_A$ are in the fundamental
and anti-fundamental representation of $\text{SU}(4)$, and $a,\dot{a}=1,2$
are the $\text{SU}(2)\times \text{SU}(2)$ little group indices. In both
\cref{eq:dimred,eq:sixd-ferm}, $q$ is a reference vector that obeys
$q\Cdot k_i\neq 0$ but otherwise is arbitrary. The final amplitudes do
not depend on the choice of $q$.

The baseline expansion brings the pure gluon and two-fermion CHY
integrand into a unified form. In order to obtain the DDM basis
numerators $N(1,\beta,n)$, we need a systematic way to expand
$\Pf(\Psi_G)$. In \cref{sec:expand}, we will discuss how this can be
done algorithmically using spanning trees.
  
\subsection{Yang-Mills-scalar}
\label{sec:yms}
Both \cref{eq:BLExp,eq:2fExp} can be interpreted as
expanding the pure-gluon/two-fermion integrand in terms of those for
single-trace YMS amplitudes. If $G$ is the set of gluons
and $(1,\tau,n)$ is the (second) color ordering of the bi-adjoint
scalars, we can write the half-integrand as~\cite{Cachazo:2014nsa}
\begin{align}\label{eq:singletrace}
\mathcal{I}^{\text{YMS}}_n\big((1,\tau,n)|G\big)=\PT(1,\tau,n)\Pf(\Psi_G)\,.
\end{align}
which corresponds to choosing the baseline factor to be
$W(1,\tau,n)=\delta_{\tau,\rho}$ in \cref{eq:BLExp}.

In fact, there exists a similar baseline expansion for the multi-trace
YMS integrand~\cite{Cachazo:2014xea},
\begin{align}\label{eq:mtrace}
  \mathcal{I}_n^{\text{YMS}}(\tau_1,\tau_2,\ldots,\tau_m|G)=\PT(\tau_1)\PT(\tau_2)\ldots\PT(\tau_m)\Pfp\Pi(\tau_1,\tau_2,\ldots,\tau_m|G)\,,
\end{align}
where $G$ denotes the set of gluons and $\{\tau_1,\ldots,\tau_m\}$ the
scalar color traces. Since in many of our future manipulations, scalar
traces are treated as a single object and have similar behavior as a
gluon, we thus define for convenience the set
\begin{align}
  \{\mathsf{t}_1,\mathsf{t}_2,\ldots,\mathsf{t}_{\mathsf{N}}\}=\{\tau_1,\tau_2,\ldots,\tau_m\}\cup G\,,\qquad \mathsf{N}=m+|G|\,,
\end{align}
where each entry $\mathsf{t}_i$ is either a scalar trace or a single
gluon. The matrix $\Pi$ is $2\mathsf{N}$ dimensional, in which the
entries are labeled by gluons and scalar traces. The explicit form of
$\Pi$ is given in \cref{sec:CHYtoAmp}. When $m=0$ and $1$, we have
\begin{align}\label{eq:mtspecial}
  \Pfp\Pi(\tau_1|G)=\Pf(\Psi_G)\,,& &\Pfp\Pi(\emptyset|G)=\Pfp(\Psi)\,,
\end{align}
such that \cref{eq:mtrace} reduces to the single trace and pure gluon
integrand respectively.

For the multitrace integrand \cref{eq:mtrace}, the baseline expansion
involves integrands with total number of gluons and trace reduced. It
applies when the end points of the baseline, chosen as $1$ and $n$, do
not belong to the same scalar trace. We can schematically write the
expansion of
$\mathcal{I}_n^{\text{YMS}}(\tau_1,\tau_2,\ldots,\tau_m|G)
=\mathcal{I}_n^{\text{YMS}}(\mathsf{t}_1,\mathsf{t}_2,\ldots,\mathsf{t}_{\mathsf{N}})$
as
\begin{align}\label{eq:BLExpMT}
\mathcal{I}_n^{\text{YMS}}(\mathsf{t}_1,\mathsf{t}_2,\ldots,\mathsf{t}_{\mathsf{N}}) &\SEq\partsum[(A,B)]{\mathsf{t}_2,\ldots,\mathsf{t}_{\mathsf{N}-1}}\Pfp\Pi(\{\mathsf{t}_1,B,\mathsf{t}_{\mathsf{N}}\},\overbrace{\alpha_1,\ldots,\alpha_s|G'\vphantom{G^{\frac{1}{2}}}}^{A})\\
  &\quad\times\PT(\alpha_1)\ldots\PT(\alpha_s)\sum_{\rho\in S_{\{\mathsf{t}_1,B,\mathsf{t_N}\}\backslash\{1,n\}}}\PT(1,\rho,n)W_{\text{MT}}(1,\rho,n)\,,\nonumber
\end{align}
where we assume that $1\in\mathsf{t}_1$ and
$n\in\mathsf{t}_{\mathsf{N}}$.  At this point, each term in the
summation of \cref{eq:BLExpMT} is a multi-trace integrand with $1$ and
$n$ in the same trace $(1,\rho,n)$, where $\rho$ is a permutations of
$\{\mathsf{t}_1,B,\mathsf{t}_{\mathsf{N}}\}$ with $1\in\mathsf{t}_1$
and $n\in\mathsf{t}_{\mathsf{N}}$ fixed at the ends. Meanwhile, the
set $A$ is given by
\begin{align}
  A=\{\mathsf{t}_{a_1},\ldots,\mathsf{t}_{a_{|A|}}\}=\{\alpha_1,\ldots,\alpha_s\}\cup G'\,,
\end{align}
where $\alpha_i$'s are scalar traces and $G'$ a subset of gluons. In
\cref{sec:rec-con-mt}, we will discuss the spanning tree algorithm to
evaluate such integrands. Before moving into more details, we
  note that \cref{eq:BLExpMT} was first derived for pure
  scalar integrands in Ref.~\cite{Du:2017gnh}. However, the
  most generic form as given here is new. We have explicitly
    checked it for varoius gluon and trace configurations against
    direct numeric evaluations of \cref{eq:CHY} up to very high
  multiplicities. It can be proved in a similar manner as
  appendix C and D of Ref.~\cite{Du:2017gnh}, but we omit such a proof
  here for conciseness.

We now introduce some essential tools to specify the baseline factor
$\BF{MT}(1,\rho,n)$ in the second row of \cref{eq:BLExpMT}. Treating
each scalar trace in $\{\mathsf{t}_1,B,\mathsf{t}_{\mathsf{N}}\}$ as a
single object, we can define a sub-ordering $(a_i,\omega_i,b_i)$ for
each $\mathsf{t}_i\in\{\mathsf{t}_1,B,\mathsf{t}_{\mathsf{N}}\}$ by
simply restricting the baseline $(1,\rho,n)$ to the elements of
$\mathsf{t}_i$,
\begin{align}
  (a_i,\omega_i,b_i)\equiv (1,\rho,n)\Big|_{\mathsf{t}_i}\,,
\end{align}
where $a_i$ and $b_i$ are respectively the first and last element of
$\mathsf{t}_i$ that appear in the baseline. If $\mathsf{t}_i$ is a
scalar trace, then $(a_i,\omega_i,b_i)$ is a permutation of
$\mathsf{t}_i$, while for $\mathsf{t}_i$ being a gluon, we simply
define $a_i=b_i=\mathsf{t}_i$ and $\omega_i=\emptyset$. Given such a
pair $(a_i,b_i)\subset\mathsf{t}_i$, we define the collection of
\emph{Kleiss-Kuijf-compatible (KK-compatible)} permutations as
\begin{align}
  \KK[\mathsf{t}_i,a_i,b_i]=\left\{\begin{array}{ccl}
                                  \left\{(a_i,\omega_i,b_i)\middle|\omega_i\in X_i\shuffle Y_i^T\right\} & & \text{if }\mathsf{t}_i=(a_i,X_i,b_i,Y_i)\text{ is a trace} \\
                                  (\mathsf{t}_i) & & \text{if }\mathsf{t}_i\text{ is a gluon}
                                \end{array}\right.\,,
\end{align}
where the set $X_i$ and $Y_i$ can be obtained by cyclically rotating
$a_i$ to the first element. They are exactly the orderings generated
by a Kleiss-Kuijf relation~\cite{Kleiss:1988ne}, and hence the
name. The sign generated by this operation
\begin{align}
  \mathfrak{sgn}^{\mathsf{t}_i}_{a_i,\omega_i,b_i}\equiv \left\{\begin{array}{ccl}
  (-1)^{|Y_i|} & \quad & \text{if }(a_i,\omega_i,b_i)\in \KK[\mathsf{t}_i,a_i,b_i] \\
  0 & &\text{otherwise}
  \end{array}\right.\,,\label{eq:sgn}
\end{align}
will be included in the baseline factor $\BF{MT}(1,\rho,n)$. Note that
for a gluon, we always have
$\mathfrak{sgn}^{\mathsf{t}_i}_{a_i,b_i}=\mathfrak{sgn}^{\mathsf{t}_i}_{\mathsf{t}_i,\mathsf{t}_i}=1$
since $|Y_i|=0$ by definition. This function is nonzero only for
KK-compatible permutations of $\mathsf{t}_i$.

In addition, we define the ordering function $\ord$ that rearranges
the baseline $(1,\rho,n)$ by comparing the first element of the
$|B|+2$ sub-orderings $(a_i,\omega_i,b_i)$: the triplet
$(a_i,\omega_i,b_i)$ precedes $(a_j,\omega_j,b_j)$ if $a_i$ precedes
$a_j$ in $(1,\rho,n)$, namely,
\begin{align}
  \ord_{(1,\rho,n)}=(\overbrace{a_0,\omega_0,b_0}^{\lambda_{0\vphantom{|B|}}},\overbrace{a_1,\omega_1,b_1}^{\lambda_{1\vphantom{|B|}}},\ldots,\overbrace{a_{|B|},\omega_{|B|},b_{|B|}}^{\lambda_{|B|}},\overbrace{a_{|B|+1},\omega_{|B|+1},b_{|B|+1}}^{\lambda_{|B|+1}})\,.
\end{align}
We can also define the \emph{coarse-grained} ordering function
\begin{align}\label{eq:Ocg}
  \ord_{(1,\rho,n)}^{\text{cg}}=(\lambda_0,\lambda_1,\ldots,\lambda_{|B|},\lambda_{|B|+1})
\end{align}
that extracts a permutation formed by the $|B|+2$ elements in
$\{\mathsf{t}_1,B,\mathsf{t}_{\mathsf{N}}\}$, which will be useful
later in section \cref{sec:rec-con-mt}. In fact, we must have
$\lambda_0=\mathsf{t}_1$ and $a_0=1$ by construction. In particular,
we will be interested in those baselines that are
\emph{$\ord$-invariant},
\begin{align}
  (1,\rho,n)=\ord_{(1,\rho,n)}\,.
\end{align}
For these cases, we must also have
$\lambda_{|B|+1}=\mathsf{t}_{\mathsf{N}}$ and $b_{|B|+1}=n$. In other
words, in $\ord$-invariant baselines every scalar trace is
\emph{consecutive} in the sense that between $a_i$ and $b_i$ there are
no elements of other traces or gluons. To characterize these
permutations, we introduces a new object
\begin{equation}
  \mathcal{B}_{(1,\rho,n)}(\lambda_i|\lambda_j) = \{ x \in \lambda_i | x \text{ before }a_j\text{ in } (1,\rho,n) \}\,,
  \label{eq:calb}
\end{equation}
that measures the ``mixing level'' between $(a_i,\omega_i,b_i)$ and $(a_j,\omega_j,b_j)$,
\begin{align}
  \mathcal{B}_{(1,\rho,n)}(\lambda_i|\lambda_j)=
  \left\{\begin{array}{ccc}
           \lambda_i & \; & \text{all of }\lambda_i\text{ come before }a_j\text{ in }(1,\rho,n) \\
           \tilde{\lambda}_i\subsetneq\lambda_i & & \text{otherwise}
         \end{array}\right.\,,
\end{align}
where $\tilde{\lambda}_i$ is a \emph{proper} subset of $\lambda_i$ and
it can be empty.  Thus the $\ord$-invariant baselines have the form
\begin{align}\label{eq:Oinv}
  (\overbrace{1,\theta_1,j_1}^{\mathsf{t}_1},\overbrace{a_1,\omega_1,b_1}^{\lambda_1},\overbrace{a_2,\omega_2,b_2}^{\lambda_2},\ldots,\overbrace{a_{|B|},\omega_{|B|},b_{|B|}}^{\lambda_{|B|}},\overbrace{i_{\mathsf{N}},\theta_{\mathsf{N}},n}^{\mathsf{t}_{\mathsf{N}}})\,,
\end{align}
and are characterized by
$\mathcal{B}_{(1,\rho,n)}(\lambda_i|\lambda_{i+1})=\lambda_i$ for
$\lambda_i\in\{\lambda_0,\lambda_1,\ldots,\lambda_{|B|}\}$. We must
have $\mathsf{t}_{\mathsf{N}}$ as the last consecutive block because
$n\in\mathsf{t}_{\mathsf{N}}$ is the endpoint of the baseline. If
there exists a $\lambda_j$ after $\mathsf{t}_{\mathsf{N}}$, then we
must have
$\mathcal{B}_{(1,\rho,n)}(\mathsf{t}_{\mathsf{N}}|\lambda_j)\subsetneq
\mathsf{t}_{\mathsf{N}}$ and thus the $\ord$-invariant condition is
violated.

With the above preparations, we now reach the punchline of the
multi-trace baseline expansion: the baseline factor
$\BF{MT}(1,\rho,n)$ is nonzero \emph{if and only if}
\begin{itemize}
\item the baseline $(1,\rho,n)$ is $\ord$-invariant: $(1,\rho,n)=\ord_{(1,\rho,n)}$;
\item the restriction of $(1,\rho,n)$ to every $\lambda_i\in\{\mathsf{t}_1,B,\mathsf{t}_{\mathsf{N}}\}$, denoted as $(a_i,\omega_i,b_i)$, is KK-compatible.
\end{itemize}
For such permutations, $\BF{MT}(1,\rho,n)=\BF{MT}(\ord_{(1,\rho,n)})$ is given by
\begin{align}
  \BF{MT}(1,\rho,n)&=\BF{MT}(\overbrace{1,\theta_1,j_1}^{\mathsf{t}_1},\overbrace{a_1,\omega_1,b_1}^{\lambda_1},\overbrace{a_2,\omega_2,b_2}^{\lambda_2},\ldots,\overbrace{a_{|B|},\omega_{|B|},b_{|B|}}^{\lambda_{|B|}},\overbrace{i_{\mathsf{N}},\theta_{\mathsf{N}},n}^{\mathsf{t}_{\mathsf{N}}})\nonumber\\
                   &=(-1)^{|B_g|}
                     \mathcal{E}^{\mathsf{t}_1}_{1,\theta_1,j_1}\Cdot \mathcal{T}_{a_1,\omega_1,b_1}^{\lambda_1}\Cdot \mathcal{T}_{a_2,\omega_2,b_2}^{\lambda_2}\Cdot\ldots\Cdot \mathcal{T}_{a_{|B|},\omega_{|B|},b_{|B|}}^{\lambda_{|B|}}\Cdot\widetilde{\mathcal{E}}^{\mathsf{t}_{\mathsf{N}}}_{i_{\mathsf{N}},\theta_{\mathsf{N}},n}\,.
\label{eq:MTbaselineComplete}
\end{align}
where $|B_g|$ is the number of gluons on the baseline, and 
\begin{subequations}
\label{eq:bsfactors}
\begin{align}
  \renewcommand{\arraystretch}{1.2}
  (\mathcal{T}^{\lambda_i}_{a_i,\omega_i,b_i})^{\mu\nu}&=\left\{\begin{array}{ccl}
                                                                  \mathfrak{sgn}_{a_i,\omega_i,b_i}^{\lambda_i}(-k_{a_i}^{\mu}k_{b_i}^{\nu}) &\; & \lambda_i\text{ is a trace and } \mathcal{B}_{(1,\rho,n)}(\lambda_i|\lambda_{i+1}) = \lambda_i  \\
                                                                  0 & & \lambda_i\text{ is a trace and } \mathcal{B}_{(1,\rho,n)}(\lambda_i|\lambda_{i+1}) \subsetneq \lambda_i \\
  f_{\lambda_i}^{\mu\nu}  & & \lambda_i\text{ is a gluon} \\
\end{array}\right.\,, \label{eq:baseline-t}\\
  (\mathcal{E}^{\lambda_i}_{a_i,\omega_i,b_i})^{\mu}&=\left\{\begin{array}{cll}
  \mathfrak{sgn}^{\lambda_i}_{a_i,\omega_i,b_i}(-k_{b_i}^{\mu}) & &\lambda_i\text{ is a trace} \\
  \pol_{\lambda_i}^{\mu} & & \text{$\lambda_i$ is a gluon}
                                                             \end{array}\right., \\
  (\widetilde{\mathcal{E}}^{\lambda_i}_{a_i,\omega_i,b_i})^{\mu}&=\left\{\begin{array}{cll}
  \mathfrak{sgn}^{\lambda_i}_{a_i,\omega_i,b_i}k_{a_i}^{\mu} & &\lambda_i\text{ is a trace} \\
  \pol_{\lambda_i}^{\mu} & & \text{$\lambda_i$ is a gluon}
  \end{array}\right..
\end{align}
\end{subequations}
Thus, the baseline factor for a generic permutation can be written as
\begin{align*}
  \BF{MT}(1,\rho,n)=(-1)^{|B_g|}\mathcal{E}^{\lambda_0}_{a_0,\omega_0,b_0}\Cdot \mathcal{T}_{a_1,\omega_1,b_1}^{\lambda_1}\Cdot \mathcal{T}_{a_2,\omega_2,b_2}^{\lambda_2}\Cdot\ldots\Cdot \mathcal{T}_{a_{|B|},\omega_{|B|},b_{|B|}}^{\lambda_{|B|}}\Cdot\widetilde{\mathcal{E}}^{\lambda_{|B|+1}}_{a_{|B|+1},\omega_{|B|+1},b_{|B|+1}}\,,
\end{align*}
where
$(\lambda_0,\lambda_1,\ldots,\lambda_{|B|},\lambda_{|B|+1})=\ord^{\text{cg}}_{(1,\rho,n)}$
is the coarse-grained baseline. The $\mathcal{B}_{(1,\rho,n)}$ and
$\mathfrak{sgn}$ in \cref{eq:bsfactors} ensure any permutations that
are not of the form of \cref{eq:MTbaselineComplete} are immediately
set to $0$.  With the help of this zeroing, the summation in the
second line of \cref{eq:BLExpMT} automatically reduces to
\begin{align}
  \sum_{\rho\in S_{\{\mathsf{t}_1,B,\mathsf{t_N}\}\backslash\{1,n\}}}\PT(1,\rho,n)W_{\text{MT}}(1,\rho,n)=\sum_{\substack{\ord\text{-invariant} \\ \text{KK-compatible}}}\PT(1,\rho,n)W_{\text{MT}}(1,\rho,n)\,.
\end{align}

Finally, we remark on the case that both the baseline endpoints $1$
and $n$ belong to the same trace, say $\tau_1$, which is left out by
\cref{eq:BLExpMT}. Starting from \cref{eq:mtrace}, we simply write
\begin{align}\label{eq:stbase}
  \mathcal{I}_n^{\text{YMS}}(\tau_1,\tau_2,\ldots,\tau_m|G)&=\PT(\tau_2)\ldots\PT(\tau_m)\Pfp\Pi(\tau_1,\tau_2,\ldots,\tau_m|G)\nonumber\\
  &\quad\times\sum_{\rho\in \KK[\tau_1,1,n]}\PT(1,\rho,n)\BF{MT}(1,\rho,n)
\end{align}
by using Kleiss-Kuijf relations. Here, we can naturally reduce the
definition of $\BF{MT}$ given in \cref{eq:MTbaselineComplete} to the
single-trace baseline,
\begin{align}
  \BF{MT}(\overbrace{1,\rho,n}^{\tau_1})=\BF{MT}(\ord_{(1,\rho,n)})=\mathfrak{sgn}^{\tau_1}_{1,\rho,n}\,.
\end{align}
We note that since the $\mathfrak{sgn}$ function is only nonzero for
KK-compatible baselines, we can trivially extend the summation range
in \cref{eq:stbase} to $S_{\tau_1\backslash\{1,n\}}$. Although this
rewriting does not reduce the total number of scalar traces and
gluons, it puts the CHY integrand into the same form as
\cref{eq:BLExpMT}, from which the half-ladder numerators can be
obtained through a unified spanning-tree algorithm given in
\cref{sec:rec-con-mt}. We note that the expansion~\eqref{eq:stbase} is first proved at the amplitude level in Ref.~\cite{Du:2017gnh}.

\section{DDM basis numerators from CHY integrands}
\label{sec:expand}
The baseline expansion is the first step of our systematic
  approach to obtain the DDM basis numerators $N(1,\beta,n)$ from CHY
  integrands. It arranges the sYM and YMS integrands into an expansion
  of YMS integrands in which the total number of gluons and traces is
  reduced. Our next step is to further expand these YMS integrands in
  terms of the DDM basis Parke-Taylor factors. This is realized
  through a \emph{spanning tree expansion}, originally
    developed in
    Refs.~\cite{Fu:2017uzt,Teng:2017tbo,Du:2017kpo,Gao:2017dek} using
    \emph{reference orderings}.  We review the original formulation
    before deriving a new approach to the construction.

\subsection{Spanning Trees and Reference Orderings}
\label{sec:ro}
We first focus on \cref{eq:BLExp,eq:2fExp}, the baseline
expansion for the pure gluon and two-fermion integrand, both of which
are a linear combination of single trace integrands
$\PT(1,\rho,n)\Pf(\Psi_G)$.
Using a different type of recursive
expansion, we can write them in terms of the single trace integrands
involving fewer gluons~\cite{Fu:2017uzt,Teng:2017tbo},
\begin{align}\label{eq:STExp}
  \PT(1,\rho,n)\Pf(\Psi_G)\SEq\sum_{(A,B)\in\Part(G\backslash l)}\Pf(\Psi_A)\sum_{\alpha\in S_B}\sum_{i=0}^{|\rho|}\mathcal{P}(l,\alpha,\rho_i)\,\mathcal{C}_{(1,\rho,n)}(\rho_i,\alpha^T,l)\,.
\end{align}
On the right hand side, we pick an arbitrary gluon $l\in G$ and sum
over all the bi-partitions of the rest. As before, both the set $A$
and $B$ can be empty. We call $\mathcal{P}(l,\alpha,\rho_i)$ a
\emph{path factor} starting from $l$ and terminating at $\rho_i$ on
the baseline (we define $\rho_0=1$), with the intermediate points a
permutation $\alpha$ of the set $B$, 
\begin{align}
\mathcal{P}(l,\alpha,\rho_i)=\pol_l\Cdot f_{\alpha_1}\Cdot f_{\alpha_2}\Cdot\ldots\Cdot f_{\alpha_{|B|}}\Cdot k_{\rho_i}\,.
\end{align}
The nomenclature will be clear in the following section in which we
represent this expansion in terms of spanning trees. Finally,
$\mathcal{C}_{(1,\rho,n)}(\rho_i,\alpha^T,l)$ is the Cayley function
of the path $(1,\rho)$ and $(\rho_i,\alpha^T,l)$~\cite{Gao:2017dek},
\begin{align}
  \mathcal{C}_{(1,\rho,n)}(\rho_i,\alpha^T,l)=\sum_{\lambda\in(\rho_{i+1},\ldots,\rho_{|\rho|})\shuffle(\alpha^T,l)}\PT(1,\rho_1,\ldots,\rho_i,\lambda,n)\,,
\end{align}
which is a linear combination of Parke-Taylor factors of length
$n{-}|A|$. The final result of this recursive expansion is the Cayley
functions of all the $(n{-}1)$-point spanning trees with a common path
$(1,\rho)$, each of which is a linear combination of DDM basis
Parke-Taylor factors~\cite{Gao:2017dek}. It is important to note that
in \cref{eq:STExp} the gauge invariance and the \crossing involving
the leg $l$ is not manifest. Since one needs to make such choices at
each step of the recursion, the final result in general will not
display any explicit gauge invariance or \crossing.

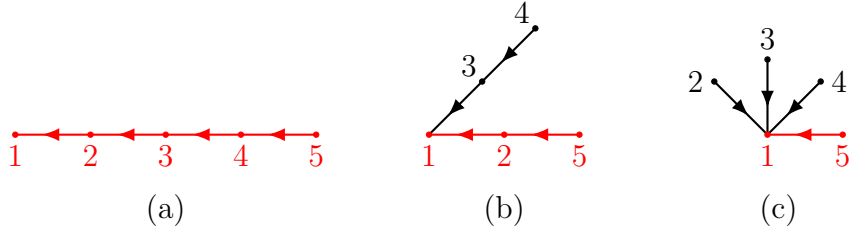
\begin{figure}[t]
  \centering
  \begin{tikzpicture}
    \draw [thick,red,dir] (1,0) -- (0,0);
    \draw [thick,red,dir] (2,0) -- (1,0);
    \draw [thick,red,dir] (3,0) -- (2,0);
    \draw [thick,red,dir] (4,0) -- (3,0);
    \filldraw [red] (0,0) circle (1pt) node[below=0pt]{$1$} (1,0) circle (1pt) node[below=0pt]{$2$} (2,0) circle (1pt) node[below=0pt]{$3$} (3,0) circle (1pt) node[below=0pt]{$4$} (4,0) circle (1pt) node[below=0pt]{$5$};
    \node at (2,-1) {(a)};
    \begin{scope}[xshift=5.5cm]
      \draw [thick,red,dir] (1,0) -- (0,0);
      \draw [thick,red,dir] (2,0) -- (1,0);
      \draw [thick,dir] (45:2) -- (45:1);
      \draw [thick,dir] (45:1) -- (45:0);
      \filldraw [red] (0,0) circle (1pt) node[below=0pt]{$1$} (1,0) circle (1pt) node[below=0pt]{$2$} (2,0) circle (1pt) node[below=0pt]{$5$};
      \filldraw (45:1) circle (1pt) node[above left=-2pt]{$3$} (45:2) circle (1pt) node[above left=-2pt]{$4$};
      \node at (1,-1) {(b)};
    \end{scope}
    \begin{scope}[xshift=10cm]
      \draw [thick,red,dir] (1,0) -- (0,0);
      \draw [thick,dir] (45:1) -- (0,0);
      \draw [thick,dir] (0,1) -- (0,0);
      \draw [thick,dir] (135:1) -- (0,0);
      \filldraw [red] (0,0) circle (1pt) node[below=0pt]{$1$} (1,0) circle (1pt) node[below=0pt]{$5$};
      \filldraw (45:1) circle (1pt) node[right=0pt]{$4$} (0,1) circle (1pt) node[above=0pt]{$3$} (135:1) circle (1pt) node[left=0pt]{$2$};
      \node at (0.146,-1) {(c)};
    \end{scope}
  \end{tikzpicture}
  \caption{Examples of trees with various $\IT$ memberships: (a) a
    tree that \emph{only} belongs to $\IT(1,2,3,4,5)$, (b) a tree that
    belongs to all $\IT(1,(2)\shuffle (3,4),5)$, (c) a tree that
    belongs to $\IT(1,\beta,5)$ for all $\beta\in S_{\{2,3,4\}}$. The
    baselines are shown in red here and in all future occurances.}
\label{fig:inc-tree-ex}
\end{figure}

With the help of \cref{eq:STExp}, one can further expand
\cref{eq:BLExp,eq:2fExp} by Parke-Taylor factors in the
DDM basis,
\begin{align}
  \mathcal{I}_n^{\text{gluon}/\text{2f}} &\SEq \partsum{2,3,\ldots,n-1} \Pf(\Psi_G) \sum_{\rho \in S_{|B|}} \PT(1, \rho , n) \BF{\text{gluon}/\text{2f}}(1,\rho,n)\nonumber\\
                & \SEq \sum_{\beta \in S_{n-2}} \PT(1, \beta, n)\ N_{\text{gluon}/\text{2f}}(1, \beta, n)\,, \label{eq:pt-nums}
\end{align}
where the coefficients $N(1,\beta,n)$ are the DDM basis
numerators. They are associated with half-ladder diagrams, and are the
master numerators for the BCJ tree relations. A given numerator
$N(1,\beta,n)$ can be constructed by assigning kinematic factors to
all the $(n-1)!$ increasing trees consistent with the label
ordering~\cite{Du:2017kpo},
\begin{equation}
  N(1, \beta, n) = \sum_{T\in \IT(1,\beta,n)} N(T) \ ,
\end{equation}
where the increasing trees are defined as
\begin{align}\label{eq:IT}
  \IT(1,\beta,n)=\left\{\text{tree }T \ \middle |\ \substack{\text{any edge }j\,\to\,i\,\in\, T \\ \text{ satisfies that $i$ is before $j$ in }(1,\beta,n)}\right\}.
\end{align}
Thus by construction $1$ is always the root of $T$ while $n$ is always
a leaf. Note that a tree $T$ can belong to several
$\IT(1,\rho,n)$. Some examples of trees and their IT membership can be
seen in \cref{fig:inc-tree-ex}.

Then, for a given increasing tree $T$,
the kinematic factor $N(T)$ is evaluated by the following algorithm~\cite{Du:2017kpo}.
\begin{alg}[Reference Order Evaluation]
  Construction of the kinematic factor for a spanning tree $T$, using a given \emph{reference ordering} $R \in S_{n{-}2}$:
  \label{al:ro}
\begin{enumerate}
\item Identify the baseline $(1,\rho,n)$, which is the path
  $n \to \rho_{|\rho|} \to \dots \to \rho_1 \to 1$ in $T$. Note that
  $\rho=\emptyset$ is allowed.
  \label{step:bf}
\item Split the rest of $T$ into \emph{ordered splitting paths}
  $\text{OS}_R(T)$ based on a \emph{reference ordering}
  $R\in S_{n-2}$:\label{step:split}
\begin{enumerate*}
\item draw a path from the first element of $R$ towards the baseline,
  which will either end on the baseline or a previously identified
  ordered splitting path.
\item move to the next element in $R$ that is not traversed yet,
  repeat the process until all vertices of $T$ are traversed.
\end{enumerate*} For example,
\begin{align}\label{eq:osexp}
  \begin{tikzpicture}[baseline={([yshift=0.1ex]current bounding box.center)},every node/.style={font=\footnotesize},scale=1]
    \draw [thick,red,dir] (1,0) -- (0,0);
    \draw [thick,red,dir] (2,0) -- (1,0);
    \draw [thick,dir] (25:2) -- (25:1);
    \draw [thick,dir] (25:1) -- (25:0);
    \filldraw [red] (0,0) circle (1pt) node[below=0pt]{$1$} (1,0) circle (1pt) node[below=0pt]{$2$} (2,0) circle (1pt) node[below=0pt]{$5$};
    \filldraw (25:1) circle (1pt) node[above left=-2pt]{$3$} (25:2) circle (1pt) node[above left=-2pt]{$4$};
  \end{tikzpicture}\xrightarrow{R=(1,2,3,4,5)}\left\{\begin{tikzpicture}[baseline={([yshift=0.1ex]current bounding box.center)},every node/.style={font=\footnotesize},scale=1]
      \draw [thick,red,dir] (1,0) -- (0,0);
      \draw [thick,red,dir] (2,0) -- (1,0);
      \filldraw [red] (0,0) circle (1pt) node[below=0pt]{$1$} (1,0) circle (1pt) node[below=0pt]{$2$} (2,0) circle (1pt) node[below=0pt]{$5$};
    \end{tikzpicture}\,,\begin{tikzpicture}[baseline={([yshift=0.1ex]current bounding box.center)},every node/.style={font=\footnotesize},scale=1]
      \draw [thick,dir] (1,0) -- (0,0);
      \filldraw (0,0) circle (1pt) node[below=0pt]{$1$} (1,0) circle (1pt) node[below=0pt]{$3$};
    \end{tikzpicture}\,,\begin{tikzpicture}[baseline={([yshift=0.1ex]current bounding box.center)},every node/.style={font=\footnotesize},scale=1]
      \draw [thick,dir] (1,0) -- (0,0);
      \filldraw (0,0) circle (1pt) node[below=0pt]{$3$} (1,0) circle (1pt) node[below=0pt]{$4$};
    \end{tikzpicture}\right\}.
\end{align}
\item Assign kinematic factors to each path
  \begin{align}
    &\text{baseline }n \to \rho_{|\rho|} \to \dots \to \rho_1 \to 1: & & W(1,\rho,n) \nonumber\\
    &l \to \alpha_1 \to \dots \to \alpha_{|\alpha|} \to i\in \text{OS}_R(T): & &\mathcal{P}(l,\alpha,i)
  \end{align}
\item $N(T)$ is the product of all of these paths,
  \begin{align}
    N(T)=W(1,\rho,n)\prod_{(l,\alpha,i)\in\text{OS}_R(T)}\mathcal{P}(l,\alpha,i)\,.
  \end{align}
  Continuing the example, the ordered splitting in \cref{eq:osexp} leads
  to
  \begin{align}
 \BF{gluon}(1,2,5)\mathcal{P}(3,1)\mathcal{P}(4,3)=\BF{gluon}(1,2,5)(\pol_3\Cdot k_1)(\pol_4\Cdot k_3)\,.
  \end{align}
\end{enumerate}
\end{alg}
The reference ordering $R$ is just a priority list of choosing the
special gluon in the recursive expansion~\eqref{eq:STExp} when
multiple choices are present. For simplicity we keep it the same for
the evaluation of all the DDM basis numerators. We note that splitting
a spanning tree into paths can be viewed as a graphic way to derive
the ordered splitting based on a reference order; the algebraic method
is first given in~\cite{Fu:2017uzt}.

While this approach is a fully constructive method of building
BCJ-compatible numerators, any \crossing is broken in each of the DDM
basis numerators, since the $\text{OS}_R$ of different color ordering
are not \crossable. This leads to one of two computation difficulties
when calculating the full integrand, either:
\begin{itemize}
\item each half-ladder numerator in the DDM basis must be computed
  individually, or
\item a given half-ladder numerator must be \emph{averaged over all
    the reference orderings} (RO-average) to make it \crossable in
  $\{2,\ldots,n-1\}$, which can then be freely relabeled to obtain the
  other half-ladder numerators.
\end{itemize}
Both approaches require an $\mathcal{O}\big((n{-}2)!\big)$ computation
(assuming that relabeling takes negligible time comparing with
computing a numerator) that must be performed after constructing a
half-ladder numerator, which itself requires an
$\mathcal{O}\big((n{-}1)!\big)$ computation, to arrive at the full DDM
basis. Thus the computational complexity for the full DDM basis
numerator is $\mathcal{O}\big[(n{-}1)!\times(n{-}2)!\big]$ for the RO
method.

\subsection{Recursive Constructions without Reference Orderings for Pure Gluon and Two-fermion Numerators}
\label{sec:rec-constr}
The computational complexity of the reference-ordering approach urges
a more efficient construction of the kinematic factors. The fact that
the \crossing among $\{2,\ldots,n-1\}$ is recovered after RO-average
hints that there should exist a new approach that solely depends on
the structure of spanning trees and does not involve reference
orderings at all. In this section, we present such an improved
algorithm.  The pure gluon and two-fermion half-ladder numerators
built through our new approach will be naturally \crossable in the
legs $\{2,\ldots,n-1\}$, removing the need for the RO-average and
reducing the complexity of construction:\footnote{This is a very
  schematic comparison, and is specifically ignoring the complexity of
  operations that are very similar between the two procedures,
  e.g. processing an individual tree.}
\begin{align}
  \mathcal{O}\big[(n{-}1)!\times(n{-}2)!\big]\;\xrightarrow{\text{new approach}}\;\mathcal{O}\big[(n{-}1)!\big] \,.
\end{align}
The final result is equivalent to the RO-average, but we only need to
evaluate each spanning tree once. We begin by considering how to build
the RO-average into simple building blocks of the increasing trees,
and then present the full, recursive approach.

The first important observation is that the baseline
$n \to \rho_{|\rho|}\to \dots \to \rho_1 \to 1$ is already independent
of the reference ordering.  As such we will carry its definition over
to our new construction.  We can then proceed to considering paths of
various lengths, and eventually more complicated trees.

The easiest paths to work with are length one. They are also
reference-ordering independent, so their contribution can be carried
over from the traditional approach,
\begin{equation}
  \begin{tikzpicture}[baseline={([yshift=0.5ex]current bounding box.center)},every node/.style={font=\footnotesize},scale=1]
    \draw [thick,dir] (0,0) -- (-1,0);
    \filldraw (0,0) circle (1pt) node[below=0pt]{$l$};
    \filldraw (-1,0) circle (1pt) node[below=0pt]{$i$};
  \end{tikzpicture}:\quad
  \phi_{l\rightarrow i}=\mathcal{P}(l,i)=\pol_l \Cdot k_i
  \label{eq:one-chain}
\end{equation}
for each such path.  Length two paths are the first object that have
reference-ordering dependence.  However, the two inequivalent
contributions dictated by the choice of reference orderings always
enter with the same weight in the average,
\begin{equation}
  \begin{tikzpicture}[baseline={([yshift=0.5ex]current bounding box.center)},every node/.style={font=\footnotesize},scale=1]
    \draw [thick,dir] (0,0) -- (-1,0);
    \draw [thick,dir] (1,0) -- (0,0);
    \filldraw (0,0) circle (1pt) node[below=0pt]{$j$};
    \filldraw (-1,0) circle (1pt) node[below=0pt]{$i$};
    \filldraw (1,0) circle (1pt) node[below=0pt]{$l$};
  \end{tikzpicture}:\;
  \phi_{l\to j \to i} = \frac{1}{2}\big[\mathcal{P}(l,j)\mathcal{P}(j,i)+\mathcal{P}(l,j,i)\big] = \frac{1}{2} \left[(\pol_l \Cdot k_j)( \pol_j \Cdot k_i) + \pol_l \Cdot f_j \Cdot k_i\right]\, .
  \label{eq:two-chain}
\end{equation}
For this particular path factor only the relative ordering between $l$
and $j$ matters: the first term $\mathcal{P}(l,j)\mathcal{P}(j,i)$ is
contributed by those reference orderings in which $j$ is before $l$
while the second term $\mathcal{P}(l,j,i)$ is from those with $l$
before $j$. Thus averaging over reference orderings assign them the
same weight.

For longer paths, it becomes more computationally difficult to
directly calculate the RO-average. Interestingly, there exists a
\emph{recursive} structure that implicitly reproduces the full
RO-average. The first important insight comes from stripping off the
$k_i$ from \cref{eq:one-chain,eq:two-chain} such that each becomes a
Lorentz vector $\vf^\mu_A$
\begin{align}
  \phi_{l \to i}&=\vf_{\{l\}}\Cdot k_i & &\longrightarrow & &\vf^{\mu}_{\{l\}} = \pol_l^\mu \label{eq:leaf-vert}\\
  \phi_{l\to j \to i} &=\vf_{\{l\to j\}}\Cdot k_i & &\longrightarrow & &\vf^{\mu}_{\{l \to j\}} = \frac{1}{2}\left[ (\pol_l \Cdot k_j) \pol_j^\mu + \pol_l^\nu f_j^{\nu \mu}\right] \label{eq:two-vert}\,.
\end{align}
The subscript on $\vf$ denotes the set of subtrees rooted on $i$.  We
can then make use of the renaming of \cref{eq:leaf-vert} in
\cref{eq:two-vert} to see the first hints of the recursion
\begin{equation}
  \vf^{\mu}_{\{l\to j\}} = \frac{1}{2} \left[ \left(\vf_{\{l\}} \Cdot k_j\right) \pol_j^\mu + \vf^{\nu}_{\{l\}} f_j^{\nu \mu} \right] \label{eq:rec-two}\,.
\end{equation}
Note that we have left the linearized field strength
  $f_j^{\nu \mu}$ unexpanded in order to more easily compare with the
  RO-average method.

While the two-vertex case is rather trivial in the
  RO-average, longer chains are not.  Luckily, the recursive structure
  continues to be simple for abitrary length chains of vertices.  To
  see this, consider the problem of finding $\vf^{\mu}_{I \to j}$,
  the partial contribution of a path
  $I_1 \to \cdots \to I_n \to j \to \cdots$ with the tensor structure
  ``downstream'' of $j$ stripped off. Inspired by
  \cref{eq:rec-two}, we make an ansatz for the recursive vertex,
  \begin{equation}
  \begin{tikzpicture}[baseline={([yshift=1ex]current bounding box.center)},every node/.style={font=\footnotesize},scale=0.8]
    \draw [thick,dir] (1,0) -- (0,0);
    \draw [thick,dir] (2,0) -- (1,0);
    \node at (2.5,0) {$\cdots$};
    \draw [thick,dir] (4,0) -- (3,0);
    \draw [thick,dir] (0,0) -- (-1,0);
    \draw [thick,dir] (-1,0) -- (-2,0);
    \filldraw (-1,0) circle (1.5pt) node[below=0pt]{$j$};
    \filldraw (4,0) circle (1.5pt) (3,0) circle (1.5pt) (2,0) circle (1.5pt) (1,0) circle (1.5pt) (0,0) circle (1.5pt);
    \draw [thick,decoration={brace,mirror,raise=0cm},decorate] (0,-0.3) -- (4,-0.3) node [pos=0.5,below=2.5pt]{$I$};
  \end{tikzpicture}\,:\quad
  \vf^{\mu}_{I \to j} = \mathfrak{a}(I)  \left(\vf_I \Cdot k_j\right)\pol_j^\mu + \mathfrak{b}(I) \vf^{\nu}_{I} f^{\nu \mu}_j\,,
\end{equation}
where $I$ is the set vertices above $j$ in the increasing tree, and
$\mathfrak{a}, \mathfrak{b}$ are combinatoric coefficients that we
will fix using induction.  First, assume that $\vf_I$ has already been
properly calculated.  Then, we want to incorporate the new information
from $j$ in such a way that we implicitly reproduce the RO average
along the path.  In the RO prescription, terms of the form
$ \left(\vf_I \Cdot k_j\right)\pol_j^\mu$ will only occur for
reference orderings in which $j$ precedes \emph{every} element of $I$.
Schematically, the ordered splitting paths are of the form
\begin{equation}
  \left(\vf_I \Cdot k_j\right)\pol_j^\mu \sim \Bigg\{
      \begin{tikzpicture}[baseline={([yshift=1ex]current bounding box.center)},every node/.style={font=\footnotesize},scale=0.8]
    \draw [thick,dir] (-2,0) -- (-3,0);
    \filldraw (-2,0) circle (1.5pt) node[below=0pt]{$j$};
    \node at (-1.5,0) {,};
    \draw [thick,dir] (0,0) -- (-1,0);
    \draw [thick,dir] (1,0) -- (0,0);
    \draw [thick,dir] (2,0) -- (1,0);
    \node at (2.5,0) {$\cdots$};
    \draw [thick,dir] (4,0) -- (3,0);
    \filldraw (-1,0) circle (1.5pt) node[below=0pt]{$j$};
    \filldraw (4,0) circle (1.5pt) (3,0) circle (1.5pt) (2,0) circle (1.5pt) (1,0) circle (1.5pt) (0,0) circle (1.5pt);
    \draw [thick,decoration={brace,mirror,raise=0cm},decorate] (0,-0.3) -- (4,-0.3) node [pos=0.5,below=2.5pt]{$I$};
  \end{tikzpicture} \Bigg\} + \cdots\,.
\end{equation}
Similarly, $\vf^{\nu}_{I} f^{\nu \mu}_j$ comes from the ROs in
which \emph{at
  least one} element of $I$ precedes $j$,
\begin{equation}
  \vf^{\nu}_{I} f^{\nu \mu}_j \sim \Bigg\{
      \begin{tikzpicture}[baseline={([yshift=0.2ex]current bounding box.center)},every node/.style={font=\footnotesize},scale=0.8]
    \draw [thick,dir] (1,0) -- (0,0);
    \node at (1.5,0) {$\cdots$};
    \draw [thick,dir] (0,0) -- (-1,0);
    \draw [thick,dir] (-1,0) -- (-2,0);
    \filldraw (-1,0) circle (1.5pt) node[below=0pt]{$j$};
    \filldraw (2,0) circle (1.5pt) node[above=0pt]{$I_i$} (1,0) circle (1.5pt) (0,0) circle (1.5pt);
    \node at (2.5,0) {,};
    \begin{scope}[xshift=4cm]
    \draw [thick,dir] (1,0) -- (0,0);
    \node at (1.5,0) {$\cdots$};
    \draw [thick,dir] (0,0) -- (-1,0);
    \filldraw (-1,0) circle (1.5pt) node[above=0pt]{$I_i$};
    \filldraw (2,0) circle (1.5pt) node[above=0pt]{$I_1$} (1,0) circle (1.5pt) (0,0) circle (1.5pt);
  \end{scope}
  \draw [thick,decoration={brace,mirror,raise=0cm},decorate] (0,-0.3) -- (6,-0.3) node [pos=0.5,below=2.5pt]{$I$};
  \end{tikzpicture}\Bigg\} + \cdots\,.
\end{equation}
A careful counting of these permutations yields
$\mathfrak{a}(I)=p|I|!$ and $\mathfrak{b}(I)=p|I||I|!$ with $|I|$ the
number of vertices in $I$, and $p$ the joint proportionality
constant. We fix this constant by demanding that, in the final result,
the term involving only a product of $\pol \Cdot k$ always carries
$+1$. Since this term is common to all the reference orderings, the
average should not change its coefficient. As a result,
$\mathfrak{a}(I)$ and $\mathfrak{b}(I)$ need to satisfy
\begin{align}
  \mathfrak{a}(I)+\mathfrak{b}(I)=1\quad\longrightarrow\quad \mathfrak{a}(I)=\frac{1}{|I|+1}\,,\quad\mathfrak{b}(I)=\frac{|I|}{|I|+1}\,.
\end{align}
Thus, the complete expression for the $\vf$ factor of an arbitrary
length path is given by the recursive definition
\begin{equation}
  \vf^{\mu}_{I \to j} = \frac{1}{|I|+1}\left[\left(\vf_I \Cdot k_j\right)\pol_j^\mu + |I| \vf^{\nu}_{I} f^{\nu \mu}_j \right] \label{eq:rec-chain}
\end{equation}
with the termination case given in \cref{eq:leaf-vert}.
Since we have constructed these terms using inductive
  matching to the RO contributions, the final result is guaranteed to
  match the RO-average calculation.

\begin{figure}[t]
  \centering
  \subfloat[Single length paths]{\centering
    \begin{tikzpicture}[baseline={([yshift=-0.ex]current bounding box.center)},scale=1]
    \draw [thick,dir] (0,0) -- (-1,0);
    \draw [thick,dir] (45:1) -- (0,0);
    \draw [thick,dir] (15:1) -- (0,0);
    \draw [thick,dir] (-45:1) -- (0,0);
    \filldraw (0,0) circle (1pt) node[below=0pt]{$j$} (45:1) circle (1pt) node[right=0pt]{$m_1$} (15:1) circle (1pt) node[right=0pt]{$m_2$} (-45:1) circle (1pt) node[right=0pt]{$m_r$};
    \filldraw (-5:1) circle (0.5pt) (-15:1) circle (0.5pt) (-25:1) circle (0.5pt);
    \draw [thick,decoration={brace,mirror,raise=0cm},decorate] (1.75,-1) -- (1.75,1) node [pos=0.5,right=2.5pt]{$M$};
    \begin{scope}[shift=(45:1cm)]
      \node at  (0,0.25)  [above=2.5pt]{$\vphantom{I_1}$};
    \end{scope}
    \begin{scope}[shift=(-45:1cm)]
      \node at  (0,-0.25) [below=2.5pt]{$\vphantom{I_r}$};
    \end{scope}
  \end{tikzpicture}\label{fig:single-length}}\quad
\subfloat[Arbitrary length chains]{\centering
  \begin{tikzpicture}[baseline={([yshift=-0.ex]current bounding box.center)}]
    \draw [thick,dir] (0,0) -- (-1,0);
    \draw [thick,dir] (45:1) -- (0,0);
    \draw [thick,dir] (15:1) -- (0,0);
    \draw [thick,dir] (-45:1) -- (0,0);
    \draw [thick,dir] (45:1) ++(1,0) -- (45:1);
    \draw [thick,dir] (45:1) ++(2,0) -- ++(-1,0);
    \draw [thick,dir] (45:1) ++(3,0) -- ++(-1,0);
    \draw [thick,dir] (15:1) ++(1,0) -- (15:1);
    \draw [thick,dir] (15:1) ++(2,0) -- ++(-1,0);
    \draw [thick,dir] (15:1) ++(3,0) -- ++(-1,0);
    \draw [thick,dir] (-45:1) ++(1,0) -- (-45:1);
    \draw [thick,dir] (-45:1) ++(2,0) -- ++(-1,0);
    \draw [thick,dir] (-45:1) ++(3,0) -- ++(-1,0);
    \filldraw (0,0) circle (1pt) node[below=0pt]{$j$} (45:1) circle (1pt) (15:1) circle (1pt) (-45:1) circle (1pt);
    \filldraw (45:1) ++(1,0) circle (1pt) (45:1) ++(2,0) circle (1pt) (45:1) ++(3,0) circle (1pt);
    \filldraw (15:1) ++(1,0) circle (1pt) (15:1) ++(2,0) circle (1pt) (15:1) ++(3,0) circle (1pt);
    \filldraw (-45:1) ++(1,0) circle (1pt) (-45:1) ++(2,0) circle (1pt) (-45:1) ++(3,0) circle (1pt);
    \filldraw (-5:1) circle (0.5pt) (-15:1) circle (0.5pt) (-25:1) circle (0.5pt);
    \draw [thick,decoration={brace,mirror,raise=0cm},decorate] (4.25,-1) -- (4.25,1) node [pos=0.5,right=2.5pt]{$M$};
    \begin{scope}[shift=(45:1cm)]
      \draw [thick,decoration={brace,mirror,raise=0cm},rotate=0,decorate] (3,0.25) -- ++(-3,0) node [pos=0.5,above=2.5pt]{$I_1$};
    \end{scope}
    \begin{scope}[shift=(-45:1cm)]
      \draw [thick,decoration={brace,mirror,raise=0cm},rotate=0,decorate] (0,-0.25) -- ++(3,0) node [pos=0.5,below=2.5pt]{$I_r$};
    \end{scope}
  \end{tikzpicture}\label{fig:arb-length}}\quad
\subfloat[Arbitrary sub-trees.]{\centering
  \begin{tikzpicture}[baseline={([yshift=-0.ex]current bounding box.center)}]
    \draw [thick,dir] (0,0) -- (-1,0);
    \draw [thick,dir] (45:1) -- (0,0);
    \draw [thick,dir] (15:1) -- (0,0);
    \draw [thick,dir] (-45:1) -- (0,0);
    \filldraw (0,0) circle (1pt) node[below=0pt]{$j$};
    \filldraw (-5:1) circle (0.5pt) (-15:1) circle (0.5pt) (-25:1) circle (0.5pt);
    \draw [thick,decoration={brace,mirror,raise=0cm},decorate] (2.25,-1) -- (2.25,1) node [pos=0.5,right=2.5pt]{$M$};
    \begin{scope}[shift=(45:1cm)]
      \draw [thick,pattern=north west lines] (0,-0.15) rectangle (1,0.15);
      \node at (0.5,0) [inner sep=0,fill=white,font=\scriptsize] {$T_1$};
      \node at  (0,0.25)  [above=2.5pt]{$\vphantom{I_1}$};
    \end{scope}
    \begin{scope}[shift=(15:1cm)]
      \draw [thick,pattern=north west lines] (0,-0.15) rectangle (1,0.15);
      \node at (0.5,0) [inner sep=0,fill=white,font=\scriptsize] {$T_2$};
    \end{scope}
    \begin{scope}[shift=(-45:1cm)]
      \draw [thick,pattern=north west lines] (0,-0.15) rectangle (1,0.15);
      \node at (0.5,0) [inner sep=0,fill=white,font=\scriptsize] {$T_r$};
      \node at  (0,-0.25) [below=2.5pt]{$\vphantom{I_r}$};
    \end{scope}
    \end{tikzpicture}\label{fig:arb-tree}}
  \caption{Merging of branches at the vertex $j$.}
  \label{fig:tree-merges}
\end{figure}
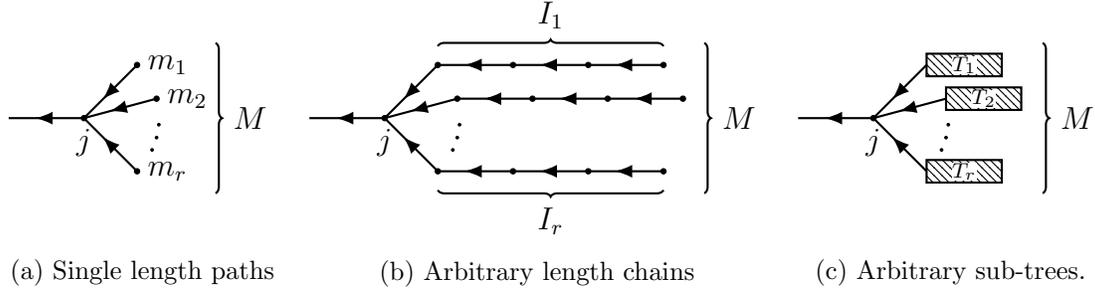

The other generalization necessary to evaluate an arbitrary tree is
the merger of multiple paths at a vertex, schematic examples of which
are shown in \cref{fig:tree-merges}. The reference-ordering approach
provides an explicit splitting of the tree at such vertices, but the
goal is to avoid such a choice.  Luckily, the recursive structure can
be continued to this case.  The simple example is the merger of two
length-one paths at a vertex. The result of RO-average is
\begin{equation}
  \begin{split}
    \begin{tikzpicture}[baseline={([yshift=-9ex]current bounding box.center)},every node/.style={font=\footnotesize},scale=0.75]
        \draw [thick,dir] (1,0) -- (0,0);
        \draw [thick,dir] (2,-.5) -- (1,0);
        \draw [thick,dir] (2,.5) -- (1,0);
        \filldraw  (0,0) circle (1pt) node[below=0pt]{$i$} (1,0) circle (1pt) node[below=0pt]{$j$} (2,.5) circle (1pt) node[above=0pt]{$l$} (2,-.5) circle (1pt) node[below=0pt]{$m$};
        \node at (2.3,0) {:};
      \end{tikzpicture}\,
  \end{split}
  \begin{split}
    \phi_{\{m,l\}\to j\to i}=\vf_{\{m,l\}\to j}\Cdot k_i =\frac{1}{3}&\Big[(\pol_m\Cdot k_j)(\pol_l\Cdot k_j)(\pol_j\Cdot k_i) \\
    &+\pol_m\Cdot f_j\Cdot k_i(\pol_l\Cdot k_j)+\pol_l\Cdot f_j\Cdot k_i(\pol_m\Cdot k_j)\Big],
  \end{split}
\end{equation}
in which there are three different contributions: one each from when
$m$ or $l$ comes before $j$ in the reference orderings, and a third
when $j$ is first among the three. Using the knowledge we have already
gained from analyzing the simpler paths in
\cref{eq:two-chain,eq:two-vert}, we can rewrite
$\vf^{\mu}_{\{m,l\}\rightarrow j}$ as
\begin{align}\label{eq:merge-two}
  \vf^{\mu}_{\{m,l\}\rightarrow j}&=\frac{1}{3}\Bigg[\pol_j^{\mu}\prod_{p\in\{m,l\}}\pol_p\Cdot k_j+\sum_{p\in\{m,l\}}\pol_p^{\nu}f_j^{\nu\mu}\prod_{\substack{q\in\{m,l\} \\ q\neq p}}\pol_q\Cdot k_j\Bigg] \nonumber\\
  &=\frac{1}{3}\Bigg[\pol_j^{\mu}\prod_{p\in\{m,l\}}\vf_{\{p\}}\Cdot k_j+\sum_{p\in\{m,l\}}\vf_{\{p\}}^{\nu}f_j^{\nu\mu}\prod_{\substack{q\in\{m,l\} \\ q\neq p}}\vf_{\{q\}}\Cdot k_j\Bigg].
\end{align}
The use of $\prod$ and $\sum$ in \cref{eq:merge-two} makes the generalization to
more incoming length-one paths straightforward: for incoming paths
from vertices $M = \{m_1, \dots, m_r\}$ we have
\begin{align}
  \text{\cref{fig:single-length}}:\;\vf^{\mu}_{M\to j}=\frac{1}{|M|+1} \Bigg[ \pol_j^{\mu} \prod_{m_a\in M} \vf_{\{m_a\}} \Cdot k_j + \sum_{m_a \in M} \vf_{\{m_a\}}^{\nu} f_j^{\nu\mu}  \prod_{\substack{m_b\in M \\ b \neq a}} \vf_{\{m_b\}} \Cdot k_j  \Bigg].
\end{align}
Finally, it is necessary to generalize each $m_a\in M$ to an arbitrary length
chain $I_a$, see \cref{fig:arb-length}. Now $M$ becomes
$M = \{I_1, \dots , I_r\}$, where each $I_a$ is a chain as in
\cref{eq:rec-chain}.  Due to the recursive structure, most of the work
was already done above in \cref{eq:rec-chain}: we can simply replace
each $\vf_{\{m_a\}}^{\mu}$ by $\vf_{I_a}^{\mu}$ and adjust the
relative coefficients,
\begin{align}
  \text{\cref{fig:arb-length}}:\;\vf^{\mu}_{M\to j}&=\mathfrak{a}\,\pol_j^{\mu} \prod_{I_a\in M} \vf_{I_a} \Cdot k_j + \sum_{I_a \in M} \mathfrak{b}_a\,\vf_{I_a}^{\nu} f_j^{\nu\mu}  \prod_{\substack{I_b\in M \\  b\neq a}} \vf_{I_b} \Cdot k_j\\
  &=\frac{1}{1+\sum_{I_a\in M}|I_a|} \Bigg[ \pol_j^{\mu} \prod_{I_a\in M} \vf_{I_a} \Cdot k_j + \sum_{I_a \in M} |I_a|\,\vf_{I_a}^{\nu} f_j^{\nu\mu}  \prod_{\substack{I_b\in M \\  b\neq a}} \vf_{I_b} \Cdot k_j  \Bigg]\,,\nonumber
\end{align}
where $\mathfrak{a}$ is the (normalized) number of reference orderings
in which $j$ is before $M$, and similarly $\mathfrak{b}_a$ is the
(normalized) number of reference orderings in which the first element
in the sub-ordering of $M\cup\{j\}$ belongs to $I_a$.

It is important to notice that the coefficients $\mathfrak{a}$ and $\mathfrak{b}_a$ depend only on the number vertices in each sub-tree but not the structure. Therefore,
this expression can be used to calculate any sub-trees
merging at one vertex, see \cref{fig:arb-tree}, where
$M=\{T_1,\ldots,T_r\}$ and each $T_a$ is an arbitrary sub-tree,
\begin{align}
  \text{\cref{fig:arb-tree}}:\;\vf^{\mu}_{M\to j}&=\frac{1}{1+\sum_{T_a\in M}|T_a|} \Bigg[ \pol_j^{\mu} \prod_{T_a\in M} \vf_{T_a} \Cdot k_j + \sum_{T_a \in M} |T_a|\,\vf_{T_a}^{\nu} f_j^{\nu\mu}  \prod_{\substack{T_b\in M \\  b\neq a}} \vf_{T_b} \Cdot k_j  \Bigg]\,,
  \label{eq:full-rec}
\end{align}
due to the recursive assumption that each $\vf_{T_a}$ has dressed its
sub-tree correctly. Thus for the most generic sub-tree structure one
only needs to correctly merge the information from each $T_a$ via the
relative normalizations, which are not sensitive
to the structure of the sub-trees.


\begin{figure}[t]
  \centering
  \begin{tikzpicture}
    \draw [red,thick,dir] (1,0) -- (0,0);
    \draw [red,thick,dir] (2,0) -- (1,0);
    \draw [red,thick,dir] (3,0) -- (2,0);
    \draw [red,thick,dir] (5,0) -- (4,0);
    \draw [red,thick,dir] (7,0) -- (6,0);
    \node at (3.5,0) [text=red] {$\cdots$};
    \node at (5.5,0) [text=red] {$\cdots$};
    \begin{scope}[xshift=1cm]
      \draw [thick,dir] (45:1) -- (0,0);
      \draw [thick,dir] (135:1) -- (0,0);
      \draw [thick,dir] (0,1) -- (0,0);
      \draw [thick,pattern=north west lines] (45:1) ++(-0.2,0) rectangle ++(0.4,1);
      \draw [thick,pattern=north west lines] (135:1) ++(-0.2,0) rectangle ++(0.4,1);
      \draw [thick,pattern=north west lines] (0,1) ++(-0.2,0) rectangle ++(0.4,1);
      \node at (0,1.5) [fill=white,inner sep=0,font=\scriptsize] {$T_a$};
      \draw [thick,decoration={brace,mirror,raise=0cm},rotate=180,decorate] (-1,-2.25) -- ++(2,0) node [pos=0.5,above=2.5pt]{$M_{\rho_1}$};
    \end{scope}
    \begin{scope}[xshift=5cm]
      \draw [thick,dir] (45:1) -- (0,0);
      \draw [thick,dir] (135:1) -- (0,0);
      \draw [thick,dir] (0,1) -- (0,0);
      \draw [thick,pattern=north west lines] (45:1) ++(-0.2,0) rectangle ++(0.4,1);
      \draw [thick,pattern=north west lines] (135:1) ++(-0.2,0) rectangle ++(0.4,1);
      \draw [thick,pattern=north west lines] (0,1) ++(-0.2,0) rectangle ++(0.4,1);
      \node at (0,1.5) [fill=white,inner sep=0,font=\scriptsize] {$T_a$};
      \draw [thick,decoration={brace,mirror,raise=0cm},rotate=180,decorate] (-1,-2.25) -- ++(2,0) node [pos=0.5,above=2.5pt]{$M_{\rho_i}$};
    \end{scope}
    \filldraw [red] (0,0) circle (1pt) node[below=0pt]{$1$} (1,0) circle (1pt) node[below=0pt]{$\rho_1$} (2,0) circle (1pt) node[below=0pt]{$\rho_2$} (3,0) circle (1pt) (4,0) circle (1pt) (5,0) circle (1pt) node[below=0pt]{$\rho_i$} (6,0) circle (1pt) circle (1pt) (7,0) circle (1pt) node[below=0pt]{$n$};
    \node at (3,1.25) {$\cdots$};
  \end{tikzpicture}
  \caption{A typical spanning tree with the baseline shown in red, where $T_a\in M_{\rho_i}$ are the sub-trees merging at $\rho_i$ on the baseline.}
  \label{fig:spanning-tree}
\end{figure}
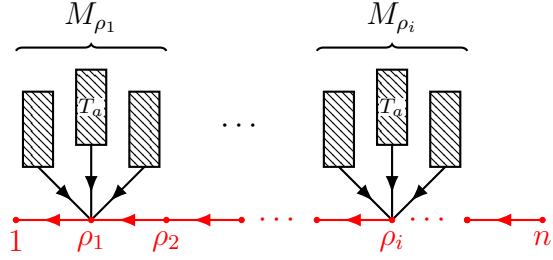

With these new kinematic dressings, we can completely get rid of
\cref{step:split} in the previous algorithm since explicit ordered
splitting computations are no longer necessary.
For completeness, our improved algorithm for calculating the numerator
contribution $N(T)$ for a given increasing tree $T$ is:
\begin{alg}[Recursive Evaluation]
  \label{al:rec}
  Construction of the kinematic factor for a spanning tree $T$
  recursively.
\begin{enumerate}
\item For the path $n \to \rho_{|\rho|}\to \dots \rho_1 \to 1$, assign
  a baseline factor $\BF{}(1,\rho,n)$. Note that $n \to 1$ and thus
  $\rho=\emptyset$ is allowed, which contributes $\BF{}(1,n)$.
\item For each vertex $\{1,\rho_1,\dots,\rho_{|\rho|}\}$ on the
  baseline, identify $M_{\rho_i}$ as the set of sub-trees merging at
  $\rho_i$, see \cref{fig:spanning-tree}. For notational convenience we define
  $\rho_0=1$.
\item Calculate $\vf$ from each sub-tree $T_a\in M_{\rho_i}$ using \cref{eq:full-rec} and assemble the result:
  \begin{equation}
    N(T) = \BF{}(1,\rho,n) \prod_{i=0}^{|\rho|}\prod_{T_a\in M_{\rho_i}} \vf_{T_a} \Cdot k_{\rho_i}\, .
    \label{eq:tree-num}
  \end{equation}
\end{enumerate}
\end{alg}
This algorithm only needs to process a single spanning tree $T$ once,
and directly generates the \crossable numerator in $\{2,\ldots,n-1\}$,
thus eliminating the $\mathcal{O}\big[(n-2)!\big]$ workload due to the
RO-average, as advertised at the beginning of this subsection. We have
explicitly checked agreement with numerators generated by RO-average
through seven points, and numerically checked that they reproduce
correct amplitudes through nine points.\footnote{The new technique can
  compute a nine-point half-ladder numerator in a few minutes, while
  the old algorithm given in \cref{sec:ro} struggles at seven
  points. Our new approach is used extensively for construction and
  checks in a parallel study of one-loop integrands from forward
  limits~\cite{Edison:2020uzf}.} We note that in our numerators $k_n$ does not appear by construction, and neither does $\pol_n\Cdot k_1$. Thus our numerators are naturally in a basis of momentum conservation.

While the procedure was designed with pure Yang-Mills trees in mind,
we can recall from \cref{eq:BLExp,eq:2fExp} that all terms that care
about $1$ and $n$ being fermions are localized to the baseline
function $\BF{2f}$.  The rest of the kinematic structure is exactly
identical to the pure Yang-Mills case.  Thus, this algorithm
\emph{also} constructs the half-ladder numerators with two fermions.

\subsection{Four-Point YM Example}
\label{sec:four-pt}
With the general principle under control, we'll now turn to an
explicit example, and demonstrate some of the functionality of the
\MMA package.  We'll construct the DDM basis numerator $N(1,2,3,4)$
using our new method, and then show that it indeed gives the correct
numerator for $N(1,3,2,4)$ via relabeling.

The first step for building the numerators is to enumerate
$\IT(1,2,3,4)$, which is implemented by the function \texttt{increasingTrees}:
\begin{equation}
  \texttt{increasingTrees[\{1,2,3,4\}]} = \left\{
  \begin{array}{cc}
    \begin{minipage}[c]{.2\textwidth}
      \begin{scaledtikzpicture}{\textwidth}
        \begin{tikzpicture}[scale=\tikzscale]
        \draw [red,thick,dir] (1,0) -- (0,0);
        \draw [red,thick,dir] (2,0) -- (1,0);
        \draw [red,thick,dir] (3,0) -- (2,0);
        \path (2,-.6) -- (2,0.6);
        \filldraw [fill=red,draw=red] (0,0) circle (3pt) node[below=0pt,text=red]{$1$} (1,0) circle (3pt) node[below=0pt,text=red]{$2$} (2,0) circle (3pt) node[below=0pt,text=red]{$3$} (3,0) circle (3pt) node[below=0pt,text=red]{$4$};
      \end{tikzpicture}
    \end{scaledtikzpicture}
  \end{minipage}%
    &%
      \begin{minipage}[c]{.17\textwidth}
      \begin{scaledtikzpicture}{\textwidth}
        \begin{tikzpicture}[scale=\tikzscale]
        \draw [red,thick,dir] (1,0) -- (0,0);
        \draw [red,thick,dir] (2,-.5) -- (1,0);
        \draw [thick,dir] (2,.5) -- (1,0);
        \filldraw [fill=red,draw=red] (0,0) circle (3pt) node[below=0pt,text=red]{$1$} (1,0) circle (3pt) node[below=0pt,text=red]{$2$}  (2,-.5) circle (3pt) node[right=0pt,text=red]{$4$};
        \filldraw (2,.5) circle (3pt) node[right=0pt]{$3$};
      \end{tikzpicture}
    \end{scaledtikzpicture}
  \end{minipage}\\
 \begin{minipage}[c]{.17\textwidth}
      \begin{scaledtikzpicture}{\textwidth}
        \begin{tikzpicture}[scale=\tikzscale]
        \draw [thick,dir] (1,.5) -- (0,0);
        \draw [red,thick,dir] (1,-.5) -- (0,0);
        \draw [red,thick,dir] (2,-.5) -- (1,-.5);
        \filldraw [fill=red,draw=red] (0,0) circle (3pt) node[below=0pt,text=red]{$1$} (1,-.5) circle (3pt) node[below=0pt,text=red]{$2$}  (2,-.5) circle (3pt) node[right=0pt,text=red]{$4$};
        \filldraw (1,.5) circle (3pt) node[above=0pt]{$3$};
      \end{tikzpicture}
    \end{scaledtikzpicture}
  \end{minipage}
    &
       \begin{minipage}[c]{.17\textwidth}
      \begin{scaledtikzpicture}{\textwidth}
        \begin{tikzpicture}[scale=\tikzscale]
        \draw [thick,dir] (1,.5) -- (0,0);
        \draw [red,thick,dir] (1,-.5) -- (0,0);
        \draw [red,thick,dir] (2,-.5) -- (1,-.5);
        \filldraw [fill=red,draw=red] (0,0) circle (3pt) node[below=0pt,text=red]{$1$} (1,-.5) circle (3pt) node[below=0pt,text=red]{$3$}  (2,-.5) circle (3pt) node[right=0pt,text=red]{$4$};
        \filldraw (1,.5) circle (3pt) node[above=0pt]{$2$};
      \end{tikzpicture}
    \end{scaledtikzpicture}
  \end{minipage}\\
    \begin{minipage}[c]{.18\textwidth}
      \begin{scaledtikzpicture}{\textwidth}
        \begin{tikzpicture}[scale=\tikzscale]
        \draw [red,thick,dir] (1,-.5) -- (0,0);
        \draw [thick,dir] (1,.5) -- (0,0);
        \draw [thick,dir] (2,.5) -- (1,.5);
        \filldraw [fill=red,draw=red] (0,0) circle (3pt) node[below=0pt,text=red]{$1$} (1,-.5) circle (3pt) node[below=0pt,text=red]{$4$};
        \filldraw  (1,.5) circle (3pt) node[above=0pt]{$2$} (2,.5) circle (3pt) node[right=0pt]{$3$};
      \end{tikzpicture}
    \end{scaledtikzpicture}
  \end{minipage}
    &
    \begin{minipage}[c]{.12\textwidth}
      \begin{scaledtikzpicture}{\textwidth}
        \begin{tikzpicture}[scale=\tikzscale]
        \draw [red,thick,dir] (1,-.5) -- (0,0);
        \draw [thick,dir] (1,.5) -- (0,0);
        \draw [thick,dir] (1,0) -- (0,0);
        \filldraw [fill=red,draw=red] (0,0) circle (3pt) node[below=0pt,text=red]{$1$} (1,-.5) circle (3pt) node[below=0pt,text=red]{$4$};
        \filldraw  (1,.5) circle (3pt) node[above=0pt]{$2$} (1,0) circle (3pt) node[right=0pt]{$3$};
      \end{tikzpicture}
    \end{scaledtikzpicture}
  \end{minipage}
  \end{array}
  \right\}\,.
\end{equation}
Then we need to dress each of the trees with their appropriate
kinematic factors.  This is handled by the function \texttt{itNumer}.
For the simplest example, we have
\begin{align}
  &\texttt{itNumer[}\begin{minipage}[c]{.14\textwidth}
      \begin{scaledtikzpicture}{\textwidth}
        \begin{tikzpicture}[scale=\tikzscale]
          \tikzset{myptr/.style={decoration={markings,mark=at position \halfway with %
                {\arrow[scale=.8,>=Latex]{>}}},postaction={decorate}}}
          \path (0,2) -- (4.5,-2);
        \draw [red,thick,myptr] (1,0) -- (0,0);
        \draw [red,thick,myptr] (2,0) -- (1,0);
        \draw [red,thick,myptr] (3,0) -- (2,0);
        \filldraw [fill=red,draw=red] (0,0) circle (3pt) node[left=0pt,text=red]{$1$} (1,0) circle (3pt) node[below=0pt,text=red]{$2$} (2,0) circle (3pt) node[below=0pt,text=red]{$3$} (3,0) circle (3pt) node[right=0pt,text=red]{$4$};
      \end{tikzpicture}
    \end{scaledtikzpicture}
  \end{minipage}\texttt{ , 4]} = \BF{gluon}(1,2,3,4)
    \,.
\end{align}
The most complicated case is actually entirely covered by \cref{eq:rec-two}:
\begin{equation}
  \texttt{itNumer[}\begin{minipage}[c]{.1\textwidth}
      \begin{scaledtikzpicture}{\textwidth}
        \begin{tikzpicture}[scale=\tikzscale]
          \tikzset{myptr/.style={decoration={markings,mark=at position \halfway with %
                {\arrow[scale=.8,>=Latex]{>}}},postaction={decorate}}}
          \path (0,2) -- (0,-2);
        \draw [red,thick,myptr] (1,-.5) -- (0,0);
        \draw [thick,myptr] (1,.5) -- (0,0);
        \draw [thick,myptr] (2,.5) -- (1,.5);
        \filldraw [fill=red,draw=red] (0,0) circle (3pt) node[left=0pt,text=red]{$1$} (1,-.5) circle (3pt) node[right=0pt,text=red]{$4$};
        \filldraw   (1,.5) circle (3pt) node[above=0pt]{$2$} (2,.5) circle (3pt) node[right=0pt]{$3$};
      \end{tikzpicture}
    \end{scaledtikzpicture}
  \end{minipage}\texttt{ , 4]}=\frac{1}{2}\left(
    (\pol_3 \Cdot k_2) (\pol_2 \Cdot k_1) + \pol_3 \Cdot f_2 \Cdot k_1\right)\BF{gluon}(1,4) \,.
\end{equation}
Combining all of the contributions together gives us
\begin{align}
  N(1,2,3,4) &= \BF{gluon}(1,2,3,4)
               + \BF{gluon}(1,2,4)\left( \pol_3 \Cdot k_2 + \pol_3 \Cdot k_1 \right)
  + \BF{gluon}(1,3,4) \pol_2 \Cdot k_1 \notag\\
  &\quad + \BF{gluon}(1,4)\left( \frac{1}{2}\left(
      (\pol_3 \Cdot k_2) (\pol_2 \Cdot k_1) + \pol_3 \Cdot f_2 \Cdot k_1\right)
    + (\pol_2 \Cdot k_1)(\pol_3 \Cdot k_1)\right) \,.
\end{align}
We can do a similar calculation to find the $(1,3,2,4)$ ordering
\begin{align}
  N(1,3,2,4) &= \BF{gluon}(1,3,2,4)
               + \BF{gluon}(1,3,4)\left( \pol_2 \Cdot k_3 + \pol_2 \Cdot k_1 \right)
  + \BF{gluon}(1,2,4) \pol_3 \Cdot k_1 \notag\\
  &\quad + \BF{gluon}(1,4)\left( \frac{1}{2}\left(
      (\pol_2 \Cdot k_3) (\pol_3 \Cdot k_1) + \pol_2 \Cdot f_3 \Cdot k_1\right)
    + (\pol_3 \Cdot k_1)(\pol_2 \Cdot k_1)\right) \,,
\end{align}
which shows that, as claimed, the two numerators are just relabelings
of each other under $(2 \leftrightarrow 3)$.  The function
\texttt{ptTreeNumer} will fully calculate an n-point tree numerator,
and relabel and dress the numerator with the appropriate Parke-Taylor factor
\begin{equation}
  \texttt{ptTreeNumer[4]} = N(1,2,3,4)\PT(1,2,3,4) + N(1,3,2,4)\PT(1,3,2,4) \,.
\end{equation}
As discussed in \cref{sec:sym}, this also builds the correct numerator
for two fermions, $N_{2\text{f}}$, by replacing
$\BF{gluon} \to \BF{2f}$.

\subsection{Recursive Constructions for Multi-trace}
\label{sec:rec-con-mt}
The recursive construction we described in \cref{sec:rec-constr,al:rec} can
be extended to also handle multi-trace YMS numerators.  We will work
out how to build numerators of the form $N_{\text{MT}}(1, \beta, n)$,
with an additional set of specified scalar traces $\tau_i$. First
using the baseline expansion discussed in \cref{sec:yms}, we can put
the legs $1$ and $n$ to the endpoints of a single scalar trace, see
\cref{eq:BLExpMT,eq:stbase}. Then we can work out the numerators
following the process first proposed in Ref.~\cite{Du:2017gnh}, in
particular, a spanning-tree algorithm given in section 10.3 (see also
appendix A of Ref.~\cite{He:2019drm}). However, this algorithm is
RO-based and thus \crossing is not manifest. In
this section, we develop additional refinements that restore
\crossing to the numerators, in line with what we did for
the pure Yang-Mills numerators in \cref{sec:rec-constr}. Similar to
the pure Yang-Mills case, the process involves two steps, constructing
spanning trees and dressing with kinematic factors. As discussed in
\cref{sec:yms}, the gluons and scalar traces are generally treated on
the same footing.

First, we discuss the structure of the increasing trees to dress, and
more generally the interplay between trace orderings and numerator
ordering. The analog of the increasing-tree ordering used in
\cref{eq:IT} is obtained by applying the coarse-grained $\ord$
function~\eqref{eq:Ocg} to the entire numerator ordering
$(1,\beta,n)$,
\begin{align}
  \ord^{\text{cg}}_{(1,\beta,n)}=(\lambda_1,\lambda_2,\ldots,\lambda_{\mathsf{N}})\in S_{\{\mathsf{t}_1,\mathsf{t}_2,\ldots,\mathsf{t}_{\mathsf{N}}\}}\,,
\end{align}
where each $\lambda_i$ represents a scalar trace or a gluon. It is an
ordering of our particle inventory
$\{\mathsf{t}_1,\mathsf{t}_2,\ldots,\mathsf{t}_{\mathsf{N}}\}=\{\tau_1,\tau_2,\ldots,\tau_m\}\cup
G$ with $\mathsf{N}=m+|G|$, where scalar traces $\tau_i$ are treated
on the same footing as a gluon in $G$. In addition, the first entry
$\lambda_1$ must be the trace that leg $1$ belongs to, or the leg $1$
itself if it is a gluon. We can then construct all the increasing
trees according to \cref{eq:IT} and write
\begin{align}
  N_{\text{MT}}(1,\beta,n)=\sum_{T\in\IT(\ord_{(1,\beta,n)}^{\text{cg}})}N_{\text{MT}}(T)\,.
\end{align}
The vertices of each tree $T$ are $\mathsf{t}_i$'s, which can either
be a scalar trace or a gluon.

Now that the increasing trees are laid out, we can begin assigning
their kinematic dressings. To evaluate $N_{\text{MT}}(T)$, the first
step is to extract the baseline from a tree $T$. Suppose
$1\in\mathsf{t}_1$ and $n\in\mathsf{t}_m$, we first extract the path
$(\mathsf{t}_1,\lambda_1,\ldots,\lambda_{|B|},\mathsf{t}_m)$ that
connects them in $T$. If $n$ is a gluon, then $\mathsf{t}_m=n$ must be
a leaf, while this is not necessary when $\mathsf{t}_m$ is a
scalar trace. Moreover, $\mathsf{t}_m$ can coincide with
$\mathsf{t}_1$. This happens when $1$ and $n$ belong to the same
scalar trace to begin with. For this case, the path consists of just a
single vertex $\mathsf{t}_1$. The baseline ordering $(1,\rho,n)$ is
then obtained by restricting the numerator color ordering
$(1,\beta,n)$ to
$(\mathsf{t}_1,\lambda_1,\ldots,\lambda_{|B|},\mathsf{t}_m)$,
\begin{align}
  (1,\rho,n)=(1,\beta,n)\Big|_{(\mathsf{t}_1,\lambda_1,\ldots,\lambda_{|B|},\mathsf{t}_m)}\,,
\end{align}
which is exactly the baseline discussed in \cref{sec:yms}. We can thus
dress it with the kinematic factor $\BF{MT}(1,\rho,n)$ defined in
\cref{eq:MTbaselineComplete}. On the other hand, given a baseline $(1,\rho,n)$, we
can recover the path
$(\mathsf{t}_1,\lambda_1,\ldots,\lambda_{|B|},\mathsf{t}_m)$, which we
call the \emph{coarse-grained baseline}, by using the coarse-grained
$\ord$ function,
\begin{align}
(\mathsf{t}_1,\lambda_1,\ldots,\lambda_{|B|},\mathsf{t}_m)=\ord_{(1,\rho,n)}^{\text{cg}}\,.
\end{align}
Eq.~(\ref{eq:ex-bl}) in the following section provides explicit
examples of this baseline identification.  Now similar to
\cref{eq:tree-num}, we can write $N_{\text{MT}}(T)$ as
\begin{align}\label{eq:NMT}
  N_{\text{MT}}(T)=\BF{MT}(1,\rho,n)\prod_{\mathsf{i}\in \ord_{(1,\rho,n)}^{\text{cg}}}\prod_{T_a\in M_{\mathsf{i}}}\mathcal{V}_{T_a}\Cdot \mathcal{K}_{\mathsf{i},T_a}\,,
\end{align}
where $M_{\mathsf{i}}$ is the set of sub-trees merging at the vertex
$\mathsf{i}$ on the coarse-grained baseline
$\ord_{(1,\rho,n)}^{\text{cg}}$. The kinematic factor associated to
each sub-tree $T_a\in M_{\mathsf{i}}$ is
$\mathcal{V}_{T_a}\Cdot \mathcal{K}_{\mathsf{i},T_a}$, where $\mathcal{K}$
will be defined below in \cref{eq:tree-k}. The vertex function $\mathcal{V}$ can be evaluated by RO-based methods~\cite{Du:2017gnh,He:2019drm}, but in the following, we
will propose an improved algorithm that automatically recovers the
\crossing.

We denote $M=\{T_1,T_2,\ldots, T_r\}$ as the set of sub-trees merging
at the vertex $\mathsf{t}_j$. By construction the vertices of our
spanning trees are labeled by
$\{\mathsf{t}_1,\ldots,\mathsf{t}_{\mathsf{N}}\}$. The recursive
structure of $\vf$ when including the vertex $\mathsf{t}_j$ takes a
form very similar to \cref{eq:full-rec},
\begin{align}\label{eq:VMT}
  \vf^{\mu}_{M\to \mathsf{t}_j}&=\frac{1}{1+\sum_{T_a\in M}|T_a|} \Bigg[ b_{\mathsf{t}_j}^{\mu} \prod_{T_a\in M} \vf_{T_a} \Cdot \mathcal{K}_{\mathsf{t}_j,T_a} + \sum_{T_a \in M} |T_a|\,\vf_{T_a}^{\nu} \mathcal{T}_{\mathsf{t}_j,T_a}^{\nu\mu}  \prod_{\substack{T_b\in M \\  b\neq a}} \vf_{T_b} \Cdot \mathcal{K}_{\mathsf{t}_j,T_a}  \Bigg]\,,
\end{align}
where $|T_a|$ counts the number of vertices in $T_a$. The combinatoric
coefficients here are the same as \cref{eq:full-rec} for pure gluon
cases since scalar traces are treated as a single object on the same
footing as gluons. The path-begin factor $b^{\mu}_{\mathsf{t}_j}$ is
given by
\begin{equation}\label{eq:tree-b}
  b_{\mathsf{t}_j}^\mu  = \begin{cases}\displaystyle
    \frac{1}{|\mathsf{t}_j|} \mathfrak{sgn}^{\mathsf{t}_j}_{a_j,\omega_j,b_j}(-k_{a_j}^\mu) & \mathsf{t}_j \text{ a trace}\\
    \pol_{\mathsf{t}_j}^\mu & \mathsf{t}_j \text{ is a gluon}
  \end{cases}\,,
\end{equation}
where $(a_j,\omega_j,b_j)$ the sub-ordering of the elements in
$\mathsf{t}_j$ in the numerator ordering $(1,\beta,n)$. The effect of
RO-average is encoded in the prefactor
$\frac{1}{|\mathsf{t}_j|}$. This prescription recovers the \crossing
within the trace $\mathsf{t}_j$. We note that in the RO approach given
in~\cite{Du:2017gnh,He:2019drm}, one has to pick the same $a_j$ across
when $\mathsf{t}_j$ appears in the path-begin factor for all the
numerator ordering, which breaks the manifest cyclicity of the scalar
traces. Next, the through factor
$\mathcal{T}_{\mathsf{t}_j,T_a}^{\mu\nu}$.  At this point, it is
useful to generalize $\mathcal{B}$ to take trees (multiple traces)
after the vertical bar via
\begin{equation}
  \mathcal{B}_{(1, \beta,n)}(\mathsf{t}_i|T_a) \equiv \bigcap_{\mathsf{t}_l\in T_a}\mathcal{B}_{(1,\beta,n)}(\mathsf{t}_j|\mathsf{t}_l) \,,
\end{equation}
that is, $\mathcal{B}_{(1,\beta,n)}(\mathsf{t}_i|T_a)$ consists of the elements
in $\mathsf{t}_i$ that come before everything in $T_a$ in the
numerator ordering $(1, \beta,n)$.  Then
$\mathcal{T}_{\mathsf{t}_j,T_a}$ is given by
\begin{align}\label{eq:tree-t}
  \mathcal{T}_{\mathsf{t}_j,T_a}^{\mu\nu}
  &=\begin{cases}
    \mathfrak{sgn}_{a_j,\omega_j,b_j}^{\mathsf{t}_j}(-k_{b_j}^{\mu}k_{a_j}^{\nu})\quad & \mathsf{t}_j\text{ is a trace and } \mathcal{B}_{(1,\beta,n)}(\mathsf{t}_j|T_a) = \mathsf{t}_j  \\
    0 &   \mathsf{t}_j\text{ is a trace and }\mathcal{B}_{(1,\beta,n)}(\mathsf{t}_j|T_a) \subsetneq \mathsf{t}_j \\
    f_{\mathsf{t}_j}^{\mu\nu}  & \mathsf{t}_j\text{ is a gluon} 
  \end{cases}.
\end{align}
Note that it takes a similar form to the $\mathcal{T}$ factor used in
the baseline factor $\BF{MT}(1,\rho,n)$, see \cref{eq:baseline-t}. The
$\mathfrak{sgn}$ function appearing in \cref{eq:tree-b,eq:tree-t}
ensures that $N_{\text{MT}}(1,\beta,n)$ is nonzero if and only if the
numerator ordering $(1,\beta,n)$ is KK-compatible to every scalar
trace. Finally, the path-end factor
$\mathcal{K}_{\mathsf{t}_j,T_a}^{\mu}$ is given by
\begin{align}
  \mathcal{K}_{\mathsf{t}_j,T_a}^{\mu}=\sum_{m\in \mathcal{B}_{(1,\beta,n)}(\mathsf{t}_j|T_a)}k_m^{\mu}\,,
  \label{eq:tree-k}
\end{align}
the total momentum from $\mathsf{t}_j$ that comes before everything in
$T_a$.  The factor $\mathcal{K}$ is actually the same as the one used
in the RO approach~\cite{Du:2017gnh,He:2019drm}. Finally, we note that the proof of \cref{eq:VMT} resembles that of section~\ref{sec:rec-constr}, which we will not repeat here.

\subsection{Multi-trace Example}
\label{sec:mt-ex}
We'll now walk through a few examples of putting the multi-trace
expansion into practice.  First, we'll calculate $N(1,2,3,4,5,6,7,8)$
where the traces are decomposed and color-ordered as
\begin{equation}
  \tau_1 = (1,8) \quad \tau_2 = (2,3) \quad \tau_3 = (4,6) \quad \tau_4 = (5,7) \,.
\end{equation}
Since $1$ and $8$ are already in the same trace, we don't need to use
the full power of the baseline recursion in this case, and can instead
focus on the kinematic factors immediately.  The increasing trees that
contribute to this numerator will have $\tau_1$ as the
baseline, and the remaining $\tau$s as vertices along the trees.
In this particle ordering, the traces are
shuffled via
\begin{equation}
  (1, \beta,8) = (1, \underbracket{2,3}_{\tau_2}, \lefteqn{\overbracket{\phantom{4,5,6}}^{\tau_3}}4,
  \lefteqn{\underbracket{\phantom{5,6,7}}_{\tau_4}}5,6,7,8) \,.
  \label{eq:ex-interlace}
\end{equation}
so the increasing trees are ordered via $\tau_2$ before $\tau_3$
before $\tau_4$.  The six increasing trees that build this numerator
can be generated using
\begin{align}
  &\texttt{increasingTrees[}\tau_1,\tau_2,\tau_3,\tau_4\texttt{]} = \notag\\
  &\qquad\qquad \begin{array}{ccc}
                   \begin{minipage}[c]{.24\textwidth}
      \begin{scaledtikzpicture}{\textwidth}
        \begin{tikzpicture}[scale=\tikzscale]
          \path (0,.6) -- (0,-.6);
        \draw [thick,dir] (1,0) -- (0,0);
        \draw [thick,dir] (2,0) -- (1,0);
        \draw [thick,dir] (3,0) -- (2,0);
        \filldraw [fill=red,draw=red] (0,0) circle (3pt) node[left=0pt,text=red]{$\tau_1$};
        \filldraw  
        (1,0) circle (3pt) node[above=0pt]{$\tau_2$}
        (2,0) circle (3pt) node[above=0pt]{$\tau_3$}
        (3,0) circle (3pt) node[right=0pt]{$\tau_4$};
      \end{tikzpicture}
    \end{scaledtikzpicture}
  \end{minipage}
    &
    \begin{minipage}[c]{.18\textwidth}
      \begin{scaledtikzpicture}{\textwidth}
        \begin{tikzpicture}[scale=\tikzscale]
        \draw [thick,dir] (1,0) -- (0,0);
        \draw [thick,dir] (2,.5) -- (1,0);
        \draw [thick,dir] (2,-.5) -- (1,0);
        \filldraw [fill=red,draw=red] (0,0) circle (3pt) node[left=0pt,text=red]{$\tau_1$};
        \filldraw  
        (1,0) circle (3pt) node[above=0pt]{$\tau_2$}
        (2,.5) circle (3pt) node[right=0pt]{$\tau_3$}
        (2,-.5) circle (3pt) node[right=0pt]{$\tau_4$};
      \end{tikzpicture}
    \end{scaledtikzpicture}
  \end{minipage}
                  &
    \begin{minipage}[c]{.12\textwidth}
      \begin{scaledtikzpicture}{\textwidth}
        \begin{tikzpicture}[scale=\tikzscale]
        \draw [thick,dir] (1,0) -- (0,0);
        \draw [thick,dir] (1,-.5) -- (0,0);
        \draw [thick,dir] (1,0) -- (0,0);
        \draw [thick,dir] (1,.5) -- (0,0);
        \filldraw [fill=red,draw=red] (0,0) circle (3pt) node[left=0pt,text=red]{$\tau_1$};
        \filldraw  
        (1,-.5) circle (3pt) node[right=0pt]{$\tau_4$}
        (1,0) circle (3pt) node[right=0pt]{$\tau_3$}
        (1,.5) circle (3pt) node[right=0pt]{$\tau_2$};
      \end{tikzpicture}
    \end{scaledtikzpicture}
  \end{minipage}\\%
    \begin{minipage}[c]{.18\textwidth}
      \begin{scaledtikzpicture}{\textwidth}
        \begin{tikzpicture}[scale=\tikzscale]
        \draw [thick,dir] (1,-.5) -- (0,0);
        \draw [thick,dir] (1,.5) -- (0,0);
        \draw [thick,dir] (2,.5) -- (1,.5);
        \filldraw [fill=red,draw=red] (0,0) circle (3pt) node[left=0pt,text=red]{$\tau_1$};
        \filldraw  
        (1,.5) circle (3pt) node[above=0pt]{$\tau_2$}
        (2,.5) circle (3pt) node[right=0pt]{$\tau_3$}
        (1,-.5) circle (3pt) node[right=0pt]{$\tau_4$};
      \end{tikzpicture}
    \end{scaledtikzpicture}
  \end{minipage}
    &
      \begin{minipage}[c]{.18\textwidth}
      \begin{scaledtikzpicture}{\textwidth}
        \begin{tikzpicture}[scale=\tikzscale]
        \draw [thick,dir] (1,-.5) -- (0,0);
        \draw [thick,dir] (1,.5) -- (0,0);
        \draw [thick,dir] (2,.5) -- (1,.5);
        \filldraw [fill=red,draw=red] (0,0) circle (3pt) node[left=0pt,text=red]{$\tau_1$};
        \filldraw  
        (1,.5) circle (3pt) node[above=0pt]{$\tau_2$}
        (2,.5) circle (3pt) node[right=0pt]{$\tau_4$}
        (1,-.5) circle (3pt) node[right=0pt]{$\tau_3$};
      \end{tikzpicture}
    \end{scaledtikzpicture}
  \end{minipage}
    &
      \begin{minipage}[c]{.18\textwidth}
      \begin{scaledtikzpicture}{\textwidth}
        \begin{tikzpicture}[scale=\tikzscale]
        \draw [thick,dir] (1,-.5) -- (0,0);
        \draw [thick,dir] (1,.5) -- (0,0);
        \draw [thick,dir] (2,.5) -- (1,.5);
        \filldraw [fill=red,draw=red] (0,0) circle (3pt) node[left=0pt,text=red]{$\tau_1$};
        \filldraw (1,.5) circle (3pt) node[above=0pt]{$\tau_3$}
        (2,.5) circle (3pt) node[right=0pt]{$\tau_4$}
        (1,-.5) circle (3pt) node[right=0pt]{$\tau_2$};
      \end{tikzpicture}
    \end{scaledtikzpicture}
  \end{minipage}
  \end{array}
\end{align}
Since $\mathcal{B}(\tau_3|\tau_4) = (4) \neq \tau_3$, the two trees
with edge $\tau_4 \to \tau_3$ are in the $\mathcal{T}^{\mu \nu}=0$
case of \cref{eq:tree-t}.  Other than that important caveat, the last
five of these trees can be evaluated by directly adapting various
examples we covered in the YM construction.  The first tree will be a
good example application of the recursive structure.  Using the
kinematic dressings from \cref{sec:rec-con-mt} in the recursion, we
have
\begin{align}
  N( & \begin{minipage}[c]{.24\textwidth}
      \begin{scaledtikzpicture}{\textwidth}
        \begin{tikzpicture}[scale=\tikzscale]
          \path (0,.6) -- (0,-.6);
        \draw [thick,dir] (1,0) -- (0,0);
        \draw [thick,dir] (2,0) -- (1,0);
        \draw [thick,dir] (3,0) -- (2,0);
        \filldraw [fill=red,draw=red] (0,0) circle (3pt) node[left=0pt,text=red]{$\tau_1$};
        \filldraw  
        (1,0) circle (3pt) node[above=0pt]{$\tau_2$}
        (2,0) circle (3pt) node[above=0pt]{$\tau_3$}
        (3,0) circle (3pt) node[right=0pt]{$\tau_4$};
      \end{tikzpicture}
    \end{scaledtikzpicture}
  \end{minipage}) = \BF{MT}(\tau_1) k_1 \Cdot \vf_{\tau_2 \tau_3 \tau_2} \notag\\
     &= -\BF{MT}(\tau_1)\frac{1}{3} k_1^\mu  \left( \frac{1}{2} k_2^\mu (k_{23}\Cdot \vf_{\tau_3 \tau_4}) + 2 k_2^\mu (k_3 \Cdot \vf_{\tau_3 \tau_4}) \right) \notag\\
     &= \BF{MT}(\tau_1) (k_1 \Cdot k_2) \frac{1}{6}\left[ (k_{23}+4 k_3)^\nu
       \frac{1}{2}\left(\frac{1}{2} k_4^\nu (k_{4}\Cdot \vf_{\tau_4}) + k_4^\nu (0)\right) \right] \notag\\
     &= -\frac{1}{48} (k_1 \Cdot k_2) (k_{4}\Cdot k_5)(k_{23}\Cdot k_4 + 4 k_3 \Cdot k_4) \,,
\end{align}
where $\BF{MT}(\tau_1) = 1$.  We do not directly provide a function to
calculate individual tree dressings for YMS due to the need to
calculate $\mathcal{B}$ and $\ord$ from the full numerator
ordering.  However, the full result from combining all of the
increasing trees can be calculated via the package using
\begin{align}
  &\texttt{multiTraceNumer[\{1,2,3,4,5,6,7,8\},\{\{1,8\},\{2,3\},\{4,6\},\{5,7\}\}]} = \notag\\
  &\qquad - \frac{1}{48} (k_1 \Cdot k_2)\left[ 2 (k_2 \Cdot k_4)(k_2 \Cdot k_5)
    + 6 (k_2 \Cdot k_5)(k_3 \Cdot k_4) + 3 (k_1 \Cdot k_5)(k_2 \Cdot k_4 + 3 k_3 \Cdot k_4) \right. \notag \\
    &\qquad\quad  + 6 (k_2 \Cdot k_4)(k_3 \Cdot k_5) + 10 (k_3 \Cdot k_4)(k_3 \Cdot k_5)
    + (k_2 \Cdot k_4)(k_4 \Cdot k_5) + 5 (k_3 \Cdot k_4)(k_4 \Cdot k_5) \notag \\
  &\qquad\quad \left. + 3 (k_1 \Cdot k_4)(2 (k_1 \Cdot k_5)
    + (k_2 \Cdot k_5) + 3(k_3 \Cdot k_5) + (k_4 \Cdot k_5))\right] \,.
\end{align}

Next, we look at a case involving gluons that demonstrates the
baseline expansion.  The goal is to calculate $N(1,2,3,4,5,6)$ with
\begin{align}
  \tau_1 = (1,2)&& \tau_2=(3,4) && G = \{5,6\} \,.
\end{align}
Since we don't
have a scalar trace of the form $(1 \dots n)$, we need to perform the
sum over pushing $\tau_2$ and $5$ into the baseline with $\tau_1$ and
$6$.  The various expansions of the baseline are given by
\begin{align}
  \begin{aligned}
    \BF{MT}(\tau_1, 6)&= k_2 \Cdot \pol_6\\
    \BF{MT}(\tau_1, \tau_2, 6)&= -(k_2 \Cdot k_3) (k_4 \Cdot \pol_6)
  \end{aligned}&&
  \begin{aligned}
   \BF{MT}(\tau_1,5,6)&= -k_2 \Cdot f_5 \Cdot \pol_6 \\
     \BF{MT}(\tau_1, \tau_2, 5, 6) &= (k_2 \Cdot k_3)(k_4 \Cdot f_5 \Cdot \pol_6)
   \end{aligned}
\end{align}
which correspond to the increasing trees
\begin{align}
  \begin{aligned}
    \BF{MT}(\tau_1,6)&:
  \begin{minipage}[c]{.11\textwidth}
      \begin{scaledtikzpicture}{\textwidth}
        \begin{tikzpicture}[scale=\tikzscale]
        \draw [red,thick,dir] (1,-.5) -- (0,0);
        \draw [thick,dir] (1,.5) -- (0,0);
        \draw [thick,dir] (2,.5) -- (1,.5);
        \path (0,2) -- (0,-2);
        \filldraw [fill=red,draw=red] (0,0) circle (3pt) node[left=0pt,text=red]{$\tau_1$}
        (1,-.5) circle (3pt) node[right=0pt,text=red]{$6$};
        \filldraw (1,.5) circle (3pt) node[above=0pt]{$\tau_2$}
        (2,.5) circle (3pt) node[right=0pt]{$5$};
      \end{tikzpicture}
    \end{scaledtikzpicture}
  \end{minipage}
    \qquad
    \begin{minipage}[c]{.11\textwidth}
      \begin{scaledtikzpicture}{\textwidth}
        \begin{tikzpicture}[scale=\tikzscale]
        \draw [red,thick,dir] (1,-.5) -- (0,0);
        \draw [thick,dir] (1,.5) -- (0,0);
        \draw [thick,dir] (1,0) -- (0,0);
        \filldraw [fill=red,draw=red] (0,0) circle (2pt) node[left=0pt,text=red]{$\tau_1$}
        (1,-.5) circle (2pt) node[below=0pt,text=red]{$6$};
        \filldraw (1,.5) circle (2pt) node[above=0pt]{$\tau_2$}
        (1,0) circle (2pt) node[right=0pt]{$5$};
      \end{tikzpicture}
    \end{scaledtikzpicture}
  \end{minipage}\\
  \BF{MT}(\tau_1,\tau_2,6)&: \begin{minipage}[c]{.11\textwidth}
      \begin{scaledtikzpicture}{\textwidth}
        \begin{tikzpicture}[scale=\tikzscale]
        \draw [red,thick,dir] (1,0) -- (0,0);
        \draw [red,thick,dir] (2,-.5) -- (1,0);
        \draw [thick,dir] (2,.5) -- (1,0);
        \filldraw [fill=red,draw=red] (0,0) circle (3pt) node[left=0pt,text=red]{$\tau_1$}
        (1,0) circle (3pt) node[below=0pt,text=red]{$\tau_2$}
        (2,-.5) circle (3pt) node[right=0pt,text=red]{$6$};
        \filldraw (2,.5) circle (3pt) node[right=0pt]{$5$};
      \end{tikzpicture}
    \end{scaledtikzpicture}
  \end{minipage} \qquad \begin{minipage}[c]{.11\textwidth}
      \begin{scaledtikzpicture}{\textwidth}
        \begin{tikzpicture}[scale=\tikzscale]
        \draw [thick,dir] (1,.5) -- (0,0);
        \draw [red,thick,dir] (1,-.5) -- (0,0);
        \draw [red,thick,dir] (2,-.5) -- (1,-.5);
        \filldraw [fill=red,draw=red] (0,0) circle (3pt) node[left=0pt,text=red]{$\tau_1$}
        (1,-.5) circle (3pt) node[below=0pt,text=red]{$\tau_2$}
        (2,-.5) circle (3pt) node[right=0pt,text=red]{$6$};
        \filldraw (1,.5) circle (3pt) node[above=0pt]{$5$};
      \end{tikzpicture}
    \end{scaledtikzpicture}
  \end{minipage}
\end{aligned}&&
                \begin{aligned}
  \BF{MT}(\tau_1,5,6)&:\begin{minipage}[c]{.11\textwidth}
      \begin{scaledtikzpicture}{\textwidth}
        \begin{tikzpicture}[scale=\tikzscale]
          \path (0,2) -- (0,-2);
        \draw [thick,dir] (1,.5) -- (0,0);
        \draw [red,thick,dir] (1,-.5) -- (0,0);
        \draw [red,thick,dir] (2,-.5) -- (1,-.5);
        \filldraw [fill=red,draw=red] (0,0) circle (3pt) node[left=0pt,text=red]{$\tau_1$}
        (1,-.5) circle (3pt) node[below=0pt,text=red]{$5$}
        (2,-.5) circle (3pt) node[right=0pt,text=red]{$6$};
        \filldraw (1,.5) circle (3pt) node[above=0pt]{$\tau_2$};
      \end{tikzpicture}
    \end{scaledtikzpicture}
  \end{minipage}\\
  \BF{MT}(\tau_1,\tau_2,5,6)&: \begin{minipage}[c]{.13\textwidth}
      \begin{scaledtikzpicture}{\textwidth}
        \begin{tikzpicture}[scale=\tikzscale]
          \setarrowscale{.8}
          \path (0,2) -- (0,-2);
        \draw [red,thick,myarr] (1,0) -- (0,0);
        \draw [red,thick,myarr] (2,0) -- (1,0);
        \draw [red,thick,myarr] (3,0) -- (2,0);
        \path (2,-.6) -- (2,0.6);
        \filldraw [fill=red,draw=red] (0,0) circle (3.2pt) node[left=0pt,text=red]{$\tau_1$}
        (1,0) circle (3.2pt) node[below=0pt,text=red]{$\tau_2$}
        (2,0) circle (3.2pt) node[below=0pt,text=red]{$5$}
        (3,0) circle (3.2pt) node[right=0pt,text=red]{$6$};
      \end{tikzpicture}
    \end{scaledtikzpicture}
  \end{minipage}
\end{aligned}
\label{eq:ex-bl}
\end{align}
We can get the combined result by again
using \texttt{multiTraceNumer}, dropping the gluons from the trace
specification
\begin{align}
  &\texttt{multiTraceNumer[\{1,2,3,4,5,6\},\{\{1,2\},\{3,4\}\}]} = \notag\\
  & \qquad \frac{1}{2}\BF{MT}(\tau_1,6) (2 \pol_5 \Cdot k_1 + 2 \pol_5 \Cdot k_2
    + \pol_5 \Cdot k_3 + 3\pol_5 \Cdot k_4)(k_1 \Cdot k_3 + k_2 \Cdot k_3)\\
  & \qquad - \BF{MT}(\tau_1,\tau_2,6) (\pol_5 \Cdot k_{1234})
    - \frac{1}{2} \BF{MT}(\tau_1,5,6)(k_1 \Cdot k_3 + k_2 \Cdot k_3) + \BF{MT}(\tau_1,\tau_2,5,6) \,.
\notag
\end{align}

\section{Introducing Masses}
\label{sec:mass}
Since the numerator construction we have described is dimensionally
agnostic (up to the choice of fermion wavefunctions), it is
straightforward to obtain numerators of minimally-coupled massive
theories through a dimensional compactification. Such a construction is
equivalent to a spontaneous symmetry
breaking~\cite{Chiodaroli:2015rdg}.
In this section, we will work with the mostly minus metric.

In particular, it is very convenient to introduce a mass to particles
$1$ and $n$ that sit at the end points of the baseline. To construct
the desired massive numerators in 4D, we start by picking special
kinematics in 6D,
\begin{equation}
  K_1 = \begin{pmatrix}k_1\\0\\-m
  \end{pmatrix},
  \qquad
  K_2 = \begin{pmatrix}k_2 \\ 0\\0
  \end{pmatrix},
  \qquad \dots \qquad
  K_n = \begin{pmatrix}k_n\\0\\m
  \end{pmatrix} \,.
  \label{eq:six-kin}
\end{equation}
where $K_i$ are 6D vectors, and $k_i$ are 4D vectors.  The 6D massless
conditions $K_1^2 = K_n^2 = 0$ impose the 4D massive conditions
$k_1^2 = k_n^2 = m^2$. For scalars, no modifications are needed as
their external wavefunctions are trivial. The gluon polarization
vectors are all kept in 4D, $\mathscr{E}_i=(\pol_i,0,0)$, and
transverse to the momentum, $K_i\Cdot\mathscr{E}_i=k_i\Cdot\pol_i=0$.

To build up the explicit embedding of massive 4D spinors into massless
6D spinors, one can follow the technique
of~\cite{Bern:2010qa,Johansson:2017bfl},
\begin{subequations}
\begin{align}
  & \rang{1_a}^A = \begin{pmatrix} \rang{1^+} & \rang{1^-}\\
    \rsqr{1^+} & \rsqr{1^-}
  \end{pmatrix}=\Big(u^{+}(k_1),u^{-}(k_1)\Big)\,, \\
  & \rsqr{1_{\dot a}}_A = \begin{pmatrix} -\lang{1^+} & \lang{1^-}\\
    -\lsqr{1^+} & \lsqr{1^-}
  \end{pmatrix}=\Big( {-}\bar{v}^{+}(k_1),\bar{v}^{-}(k_1)\Big)\,, \\
  & \rang{n_a}^A = \begin{pmatrix} -\rang{n^+} & \rang{n^-}\\
    \rsqr{n^+} & -\rsqr{n^-}
  \end{pmatrix}=\Big(v^{+}(k_n),-v^{-}(k_n)\Big)\,, \\
  & \rsqr{n_{\dot a}}_A = \begin{pmatrix} \lang{n^+} & \lang{n^-}\\
                     -\lsqr{n^+} & -\lsqr{n^-}
  \end{pmatrix}=\Big( {-}\bar{u}^{+}(k_n) , \bar{u}^{-}(k_n) \Big)\,,
\end{align}
\end{subequations}
where $|i^{\pm}\rangle$ and $|i^{\pm}]$ are massive spinor helicity
variables~\cite{Arkani-Hamed:2017jhn}, and $u$ and $v$ are usual
momentum-space Dirac spinors. The explicit representation of 6D
spinors depends on the choice of 6D gamma matrices, and here we use
the one given in the appendix A of
Ref.~\cite{Cheung:2009dc}.\footnote{This is our motivation for putting
  the mass into the sixth slot instead of the fifth in
  \cref{eq:six-kin}, as this choice will lead to a simpler embedding
  of fermion wavefunctions.} Our definition of Dirac spinors in terms
of massive spinor helicity variables follows
Ref.~\cite{Ochirov:2018uyq}, where one can also find explicit formulas
for these quantities.

We note that all the formal expressions for our numerators are
unchanged under the uplift~\eqref{eq:six-kin}, except that one needs
to use massive kinematics when evaluating. In fact, the only way to
introduce an explicit mass term is via
$K_1 \Cdot K_n = k_1 \Cdot k_n + m^2$ while
$K_i\Cdot K_j=k_i\Cdot k_j$ for all the rest. Our numerators are
constructed to only contain the wavefunction of particle $n$ but not
the momentum, such that the product $K_1 \Cdot K_n$ will not naturally
appear unless one deliberately swaps it in via momentum conservation.

\section{Conclusion and Outlook}
In this paper we presented a review of the current state-of-the-art
for calculating YM tree amplitudes in the CHY formalism.  We
identified a computational complexity that results from the
requirement of the previous methods, summarized in \cref{al:ro}, to define an \emph{ordered
  splitting of paths} in order to build kinematic factors, which in
turn builds numerators that are not \crossable.  To overcome this
complexity, we developed a new method of assigning kinematic factors
to increasing trees that incorporates all contributions from different
\emph{ordered splittings} implicitly, rather than via an explicit
after-the-fact average. This new method, summarized in
  \cref{al:rec}, is built around a recursive processing of the
  underlying spanning trees, allowing the required kinematic factors
  to be determined at each vertex by using a small amount of
  information from the previous steps in the recursion. One of the
  benefits of the new approach is the
restoration of \crossing at every step in the calculation,
  without needing to process the same tree multiple times.  Our new
numerator construction is now \emph{only} a function of external
particle ordering, and has no dependence on reference orderings. The
new method was also straightforward to generalize to include full
multi-trace YMS numerators or two-fermion numerators.  Since the
numerators we construct are in the DDM basis, they can be
double-copied immediately to yield the corresponding numerators for
Einstein-Yang-Mills.

A related benefit of the new method is a signficant
  improvement in the accesible multiplicity of tree amplitudes.  While
  the RO based methods require processing all of the spanning trees
  again for each numerator in \cref{eq:ym-amp,eq:grav-amp}, the new
  algorithm can simply relabel a prototype numerator.  As such our
new method is efficient enough that we can easily calculate 8 point YM
trees, where two of the external legs can be swapped out for fermions
or scalars instead; we have been able to calculate individual
\crossable tree numerators up to 10 points without significant
difficulty.  For 7 and fewer external particles, we have
  explicitly compared our results with the RO average.  At 8 points
  and beyond, we have verified our results by comparison
  to a direct numerical calculation of \cref{eq:CHY} on certain
  solutions of the scattering equations. This ability to calculate
high-point trees with super-partners has been incredibly useful in the
study of forward limits, which we leave to its own paper
\cite{Edison:2020uzf}.

The fermionic baseline functions are the only piece of the calculation
that depend on the dimension of interest. Thus, we described one
method of using dimensional compactification to introduce a masses to
the 2 singled out particles labeled $1$ and $n$.  We hope that the
combination of double-copy compatibility, potentially massive
particles, and calculation efficiency will be of interest to the
growing application of amplitudes techniques to classical gravity, for
instance as input into the unitarity cuts approach of constructing
relevant loop
amplitudes~\cite{Cheung:2018wkq,Guevara:2018wpp,Bern:2019nnu,Guevara:2019fsj,Johansson:2019dnu,Bern:2019crd}.

Finally, we provide a \MMA package the implements the recursive
numerator construction.  Appendix \ref{sec:package} and the examples in
\cref{sec:four-pt,sec:mt-ex} have more details about the package and
its usage.

\section*{Acknowledgments}
We would like to thank Song He and Oliver Schlotterer for sharing
unpublished notes related to the fermionic expansion
\cref{eq:2fExp,eq:WF1,eq:WF2}.  We would additionally like to thank
Song He, Henrik Johansson, and Oliver Schlotterer for comments on the
manuscript, and Maor Ben-Shahar and Gregor K{\"a}lin for testing the
\MMA package.  AE is supported by the Knut and Alice Wallenberg
Foundation under KAW 2018.0116, {\it From Scattering Amplitudes to
  Gravitational Waves.}  FT is supported in part by the Knut and Alice
Wallenberg Foundation under grant KAW 2013.0235, and the Ragnar
S\"{o}derberg Foundation (Swedish Foundations’ Starting
Grant). Bibliography produced in part with the help of
fill\TeX{}~\cite{2017JOSS....2..222G}.

\appendix
\section{CHY integrands and scattering amplitudes}
\label{sec:CHYtoAmp}
In this appendix, we provide the explicit forms of the CHY integrands
used in the main text and the normalizations used to match Feynman
diagram results. Throughout the paper, we use the following definition
for the Pfaffian of a $(2n)\times(2n)$ antisymmetric matrix $X$,
\begin{align}\label{eq:pfaffian}
\Pf(X)\equiv(-1)^{\frac{n(n+1)}{2}}\frac{1}{2^n n!}\sum_{\sigma\in
  S_{2n}}\text{sign}(\sigma)\prod_{i=1}^{n}X_{\sigma_{2i-1}\sigma_{2i}}\,.
\end{align}
The sign factor $(-1)^{\frac{n(n+1)}{2}}$ is not present in the usual
definition of a Pfaffian~\cite{pfaffian}, but convenient for our
purposes.

In the pure gluon integrand~\eqref{eq:gluon}, the reduced Pfaffian is defined as
\begin{align}
  \Pfp(\Psi)\equiv\frac{(-1)^{i+j+n-1}}{\sigma_{ij}}\Pf(\Psi_{ij}^{ij})\,,& & 1\leqslant i<j\leqslant n.
\end{align}
where the $2n\times 2n$ antisymmetric matrix $\Psi$ is given in the block form,
\begin{align}
  \Psi=\left(\begin{array}{cc}
  \mathsf{A} & -\mathsf{C}^T \\
  \mathsf{C} & \mathsf{B}
  \end{array}\right).
\end{align}
The three $n\times n$ blocks $\mathsf{A}$, $\mathsf{B}$ and $\mathsf{C}$ are given by
\begin{align}\label{eq:ABC}
  & \mathsf{A}_{ij}=\frac{k_i\Cdot k_j}{\sigma_{ij}}\,, & & \mathsf{B}_{ij}=\frac{\pol_i\Cdot\pol_j}{\sigma_{ij}}\,, & & \mathsf{C}_{ij}=\frac{\pol_i\Cdot k_j}{\sigma_{ij}}\,, \nonumber\\
  & \mathsf{A}_{ii}=0\,, & & \mathsf{B}_{ii}=0\,, & & \mathsf{C}_{ii}=-\sum_{\substack{j=1 \\ j\neq i}}^n \mathsf{C}_{ij}\,.
\end{align}
The polarization vectors are normalized as $\pol_i\Cdot\pol_i^{*}=1$. The matrix $\Psi_{ij}^{ij}$ is obtained from $\Psi$ by deleting the
$i$-th and $j$-th row and column. On the support of the scattering
equations, $\Pfp(\Psi)$ is independent of the choice of $i$ and
$j$. The $\Psi_G$ that appears in the baseline
expansion~\eqref{eq:BLExp} and the single-trace
integrand~\eqref{eq:singletrace} is a $2|G|\times 2|G|$ antisymmetric
submatrix of $\Psi$, obtained by restricting the rows and columns in
the $\mathsf{A}$, $\mathsf{B}$ and $\mathsf{C}$ blocks to the set $G$,
\begin{align}\label{eq:PsiG}
  \Psi_G=\left(\begin{array}{cc}
    \mathsf{A}_G & -(\mathsf{C}_G)^T \\
    \mathsf{C}_G & \mathsf{B}_G
  \end{array}\right),
\end{align}
where $(\mathsf{X}_G)_{ij}=\mathsf{X}_{ij}$ for $\mathsf{X}=\mathsf{A},\mathsf{B},\mathsf{C}$ and $i,j\in G$. 

The multi-trace integrand $\mathcal{I}_n(\beta_1,\ldots,\beta_m|G)$ in
\cref{eq:mtrace} contains the reduced Pfaffian of the
$2(m+|G|)\times 2(m+|G|)$ antisymmetric matrix $\Pi$, which can also
be written in a block form,
\begin{align}
  \Pi(\tau_1,\tau_2,\ldots,\tau_m|G)=\left(\begin{array}{cccc}
    \mathsf{A}_G & \mathsf{A}_{G,\text{tr}} & -(\mathsf{C}_G)^T & -(\mathsf{C}_{\text{tr},G})^T \\
    \mathsf{A}_{\text{tr},G} & \mathsf{A}_{\text{tr}} & -(\mathsf{C}_{G,\text{tr}})^T & -(\mathsf{C}_{\text{tr}})^T \\
    \mathsf{C}_G & \mathsf{C}_{G,\text{tr}} & \mathsf{B}_G & \mathsf{B}_{G,\text{tr}} \\
    \mathsf{C}_{\text{tr},G} & \mathsf{C}_{\text{tr}} & \mathsf{B}_{\text{tr},G} & \mathsf{B}_{\text{tr}}
    \end{array}\right),
\end{align}
where the blocks are defined as follows,
\begin{itemize}
\item The $|G|\times|G|$ matrix $\mathsf{A}_G$, $\mathsf{B}_G$ and $\mathsf{C}_G$ are just those appeared in $\Psi_G$ of \cref{eq:PsiG}.
\item The $m\times|G|$ matrix $\mathsf{A}_{\text{tr},G}$, $\mathsf{B}_{\text{tr},G}$ and $\mathsf{C}_{\text{tr},G}$ are given by
  \begin{align}
    (\mathsf{A}_{\text{tr},G})_{ib}=\sum_{c\in\tau_i}\frac{k_c\Cdot k_b}{\sigma_{cb}}\,,& &(\mathsf{B}_{\text{tr},G})_{ib}=\sum_{c\in\tau_i}\frac{\sigma_ck_c\Cdot\pol_b}{\sigma_{cb}}\,,& &(\mathsf{C}_{\text{tr},G})_{ib}=\sum_{c\in\tau_i}\frac{\sigma_ck_c\Cdot k_b}{\sigma_{cb}}\,.
  \end{align}
\item The $|G|\times m$ matrix $\mathsf{A}_{G,\text{tr}}$,
  $\mathsf{B}_{G,\text{tr}}$ and $\mathsf{C}_{G,\text{tr}}$ are given
  by
  \begin{align}
    \mathsf{A}_{G,\text{tr}}=-(\mathsf{A}_{\text{tr},G})^T\,,& &\mathsf{B}_{G,\text{tr}}=-(\mathsf{B}_{\text{tr},G})^T\,,& &(\mathsf{C}_{G,\text{tr}})_{ai}=\sum_{c\in\tau_i}\frac{\pol_a\Cdot k_c}{\sigma_{ac}}\,.
  \end{align}
  \item The $m\times m$ matrix $\mathsf{A}_{\text{tr}}$, $\mathsf{B}_{\text{tr}}$ and $\mathsf{C}_{\text{tr}}$ are given by
    \begin{align}
      & (\mathsf{A}_{\text{tr}})_{ij}=\sum_{\substack{c\in\tau_i \\ d\in\tau_j}}\frac{k_c\Cdot k_d}{\sigma_{cd}}\,, & & (\mathsf{B}_{\text{tr}})_{ij}=\sum_{\substack{c\in\tau_i \\ d\in\tau_j}}\frac{\sigma_c\sigma_dk_c\Cdot k_d}{\sigma_{cd}}\,, & & (\mathsf{C}_{\text{tr}})_{ij}=\sum_{\substack{c\in\tau_i \\ d\in\tau_j}}\frac{\sigma_ck_c\Cdot k_d}{\sigma_{cd}}\,, \nonumber\\
  & (\mathsf{A}_{\text{tr}})_{ii}=0\,, & & (\mathsf{B}_{\text{tr}})_{ii}=0\,, & & (\mathsf{C}_{\text{tr}})_{ii}=\frac{1}{2}(k_{\tau_i})^2\,,
    \end{align}
    where $k_{\tau_i}=\sum_{c\in\tau_i}k_c$ is the total momentum of the trace $\tau_i$.
\end{itemize}
The reduced Pfaffian $\Pfp(\Pi)$ is defined as
\begin{align}
  \Pfp(\Pi)&=(-1)^{i+j}\Pf(\Pi_{|G|+i,2|G|+m+j}^{|G|+i,2|G|+m+j}) & &1\leqslant i,j\leqslant m \nonumber\\
  &=(-1)^{|G|+m-1}\frac{(-1)^{a+b}}{\sigma_{ab}}\Pf(\Pi_{ab}^{ab}) & &a,b\in G\text{ and }a<b\,.
\end{align}
The two expressions are equivalent on the support of the scattering
equations when both gluons and traces are present, while the first
line applies for the pure scalar case and the second line applies for
the pure gluon case. For these two special cases, one can then easily
check that \cref{eq:mtspecial} holds.

To explicitly calculate amplitudes from the CHY formalism, one needs
to perform the moduli space integral in \cref{eq:CHY}. It turns the
delta functions into a Jacobian and results in a summation over the
$(n-3)!$ solutions of the scattering equations. Starting with the
gauge fixing~\eqref{eq:wsgf}, we get
\begin{align}
  A_n(1,2,\ldots,n)=\mathscr{N}\sum_{\text{solutions}}\frac{(\sigma_{1,n-1}\sigma_{n-1,n}\sigma_{n,1})^2}{\det(\Phi_{1,n-1,n}^{1,n-1,n})}\PT(1,2,\ldots,n)\,\mathcal{I}_n(\beta_1,\ldots,\beta_m|G)\,,
\end{align}
where the matrix $\Phi$ is given by
\begin{align}
  \Phi_{ij}=\frac{k_i\Cdot k_j}{\sigma_{ij}^2}\,,& &\Phi_{ii}=-\sum_{\substack{j=1 \\ j\neq i}}^n\Phi_{ij}\,.
\end{align}
To obtain the Jacobian $\det(\Phi_{1,n-1,n}^{1,n-1,n})$, one needs to
delete three rows and columns from $\Phi$ following the gauge
fixing. The normalization $\mathscr{N}$ is given by
\begin{align}
  \mathscr{N}=(-2i)\left(\frac{g}{\sqrt{2}}\right)^{|G|+2m-2}\left(-\frac{\lambda}{4}\right)^{|S|-2m}(-1)^m\,,
  \label{eq:expl-norm}
\end{align}
where $|S|=n-|G|$ is the total number of scalars. This normalization
is fixed by comparing with the color-ordered amplitudes from the
YMS Lagrangian \cite{Chiodaroli:2014xia}
\begin{align}
  \mathcal{L}=-\frac{1}{4}(F^a_{\mu\nu})^2+\frac{1}{2}(D_{\mu}\phi^{aA})^2+\frac{\lambda}{3!}f^{abc}\hat{f}^{ABC}\phi^{aA}\phi^{bB}\phi^{cC}-\frac{g^2}{4}f^{ace}f^{ebd}\phi^{aA}\phi^{bA}\phi^{cB}\phi^{dB}\,,
\end{align}
with the gauge coupling $g$ and $\phi^3$ coupling $\lambda$. The
scalar $\phi^{aA}$ is in the bi-adjoint representation of the gauge
group and another global symmetry group, whose structure constants are
$f^{abc}$ and $\hat{f}^{ABC}$ respectively. In particular, the
normalization for the pure gluon cases is
\begin{align}
  \mathscr{N}=(-2i)\left(\frac{g}{\sqrt{2}}\right)^{n-2}\,,
\end{align}
which is shared by the two-fermion integrand~\eqref{eq:2fExp} since
the two amplitudes are related by a SUSY Ward identity. The
outstanding factor $2$ can be absorbed into the definition of the
reduced Pfaffian, as pointed out in the original CHY
paper~\cite{Cachazo:2013hca}, while the rest simply originate from
rescaling color factors. More specifically, we define partial
amplitudes in the color trace bases with the normalization
$\text{Tr}(T^aT^b)=\delta^{ab}$ and
$\text{Tr}(T^AT^B)=\delta^{AB}$. Both traces are taken in the
fundamental representation of the gauge group.

\section{\MMA Package}
\label{sec:package}
This appendix gives a more in-depth description for the functions
provided by our \MMA package, \texttt{IncreasingTrees.m}.  The package
can be found by following this link to our GitLab\footnote{For those averse to clicking links, \url{https://gitlab.com/aedison/increasingtrees}}:
\begin{center}
  \href{https://gitlab.com/aedison/increasingtrees}{IncreasingTrees GitLab}.
\end{center}
There are no external dependencies for the package, although it does
rely on the \texttt{Graph} functionality of \MMA\!\!, and thus may
depend on the version in use.  We have tested the package on \MMA
versions \texttt{11.2} and \texttt{12.0}.

The package defines a set of objects from which it builds the
kinematic numerators, as well as defining construction routines to
build the various numerators.  It localizes all variables within
either its own context \texttt{IncreasingTrees\`} or its private
context \texttt{IncreasingTrees\`{}Private\`}, and thus is generally
safe to load along other packages.  However, it does provide
definitions for the single-character variables \texttt{k,e,W,d,F}, which
may shadow definitions from other packages.

\subsection{Constituent objects}
The following objects are used to represent the numerators:

\defn{k}{\vardefHH{i}{}}{Momentum vector for particle(s) \varN{i}.
  Multi-label \texttt{k} are automatically converted to sums:
  \texttt{k[1,2]}$=k_1 + k_2$.}

\defn{e}{\vardefH{i}{}}{Polarization vector for particle \varN{i}.}

\defn{d}{\vardefH{x}{},\vardefH{y}{}}{Lorentz dot product between two
  vectors.  Orderless, bilinear, and enforces transversality and
  onshellness of \texttt{k} and \texttt{e}.}

\defn{W}{\vardefHH{}{}}{Generic baseline factor.  Can be swapped out
  directly using replacement rules for \texttt{WScalar},
  \texttt{WGluon}, \texttt{WFermion2}, which are all described in
  \cref{sec:chy}.  \texttt{WL} and \texttt{WR} are additionally
  defined to allow for differentiating theories in the double copy.}

\defnn{WFermion1}{\vardefH{i}{}}{\vardefHH{}{}}{Baseline factor for
  the $(-\frac{1}{2},-\frac{1}{2})$ ghost picture fermions in
  \cref{eq:WF1}.  \varN{i} is the gluon in ghost picture $-1$.  For
  example, replacing \texttt{W}$\to$\texttt{WFermion1[2]} in a
  numerator will apply this baseline with gluon $2$ in the alternate
  ghost picture.}

\defn{spinChain}{\vardefHH{}{}}{Abstract spin index contraction
  generated by \texttt{WFermion}s.}

\defn{$\chi$}{\vardefH{i}{}}{Fermionic wavefunction placeholder for particle \varN{i}.}

\defn{$\xi$}{\vardefH{i}{}}{Ghost picture $-\frac{3}{2}$ fermionic
  wavefunction for particle \varN{i}.}

\defn{F}{\vardefH{i}{}}{Linearized gluon field-strength.  Only appears
  explicitly in the fermion baseline functions
  \cref{eq:WF1,eq:WF2}. We give explicit $\gamma$ contractions in
  \cref{eq:tr-f}.}

\defn{PT}{\vardefHH{}{}}{Abstract Parke-Taylor factor used as part of
  representing (half-)integrands.}

\defn{cf}{\vardefHH{}{}}{Abstract color factor for color-dressed
  amplitude construction, following \cref{eq:ym-amp}.}

\subsection{Construction functions}
The package exposes the following functions used in assembling
half-ladder numerators:

\defn{increasingTrees}{\vardefH{labels}{List}}{Constructs all ordered
  trees where the vertex labels are taken from \varN{labels} and the
  ordering is consistent with the ordering of \varN{labels}.  Thus,
  each tree spans the vertices and is only composed of
  \texttt{DirectedEdges} of the form \varN{labels}[[\,j]] $\to$
  \varN{labels}[[i]] with $i<j$.
}

\defn{itNumer}{\vardefH{g}{Graph},\ \vardef{maxV}}{ Processes \varN{g}
  to extract the baseline information and runs the recursive
  construction (detailed in \cref{sec:rec-constr}) starting from each vertex along the baseline to
  generate a kinematic polynomial. \varN{maxV} specifies which vertex
  should be treated as the end of the baseline.
}

\defn{ptTreeNumer}{\vardefH{points}{Integer},\
  \vardef{onlyFuncQ}{\color{varcolor}\texttt{:False}}}{ This function
  has a different return scheme based on whether \varN{onlyFuncQ} is
  \texttt{True} or \texttt{False}:
  \begin{itemize}
    \item \texttt{True}:
    generates a \texttt{Function} with \varN{points} slots that in
    turn produces a half-ladder numerator with the
    specified external labels.
  \item \texttt{False}(default): generates \emph{all} half-ladder
    numerators with \varN{points} number of external particles,
    dressed with appropriate PT factors.  Explicitly uses the
    \crossing of labels $(2, \dots, \varN{points}{-}1)$.
\end{itemize}

}

\defn{multiTraceNumer}{\vardefH{labels}{List},\
  \vardefH{traces}{List}}{ Computes the kinematic numerator for a mix
  of scalars and gluons carrying labels specified and ordered in
  \varN{labels}, grouped into scalar traces specified by
  \varN{traces}.  Handles the construction and insertion of the
  baseline factor directly. Traces should be given as a \texttt{List}
  of \texttt{List}s, where each sublist specifies a color-ordered
  trace of scalar particles.  Gluons can either be specified as length
  1 sublists in \varN{traces}, or omitted from the trace specification
  entirely.  Expects each particle to be labeled the same way in
  \varN{labels} and \varN{traces}.  }

\defn{multiTraceIntegrand}{\vardefH{points}{Integer},\
  \vardefH{traces}{List}}{Computes the explicit permutation sum of
  \texttt{multiTraceNumer} over the permutations of labels ranging
  from $(2, \dots, $\varN{points}$-1)$, using the trace configurations
  given in \varN{traces}, and dressed with appropriate PT factors. The
  particle labels in \varN{traces} should be given as integers
  $1, \dots , $\varN{points}.}

\noindent Unfortunately, since the symmetries of the multi-trace
numerators are significantly more nuanced than in the case of only
gluons (or gluons with two scalars or two fermions), we don't make
direct use of them in \texttt{multiTraceIntegrand} like we do in
\texttt{ptTreeNumer}.  As such, while \texttt{multiTraceIntegrand} and
\texttt{ptTreeNumer} produce identical results for pure gluons, we
strongly recommend to use \texttt{ptTreeNumer} in these situations,
especially for larger particle number.

In addition to the half-ladder numerator constructors, we also provide
the tools to assemble the numerators into amplitudes.  However, these
computations run into the $\mathcal{O}((n{-}2)!^2)$ complexity of
calculating the full matrtix of $m(\alpha|\beta)$.

\defn{mab}{\vardefH{$\alpha$}{List},\vardefH{$\beta$}{List}}{%
  Computes the biadjoint scalar amplitude for orderings
  \varN{$\alpha$} and \varN{$\beta$}.  The orderings should be in the
  same KK basis: $\alpha_1 = \beta_1$ and $\alpha_n = \beta_n$.  The
  momentum $k_{\alpha_n}$ is implicitly removed via momentum
  conservation. Uses the algorithm proposed by
  Ref.~\cite{Mafra:2016ltu}.}

\defn{Aamp}{\vardefH{labels}{List}}{Constructs the Yang-Mills tree
  amplitude according to \cref{eq:ym-amp}, where the external
  particle labels are drawn from \varN{labels}.  Includes the
  explicit normalization factor \cref{eq:expl-norm}. The baseline
  factors \texttt{W[\dots]} are left unevaluated to allow for choice of
  special particles.}

\defn{Aamp}{\vardefH{n}{Integer}}{Constructs the YM tree amplitude
  using $1,\dots, n$ as the labels.}

\defn{Mamp}{\vardefH{labels}{List}}{Constructs the gravity tree
  amplitude according to \cref{eq:grav-amp}, where the external
  particle labels are drawn from \varN{labels}.  The baseline factors
  for the two copies are left unevaluated, with seperate labels \texttt{WL}
  and \texttt{WR} to allow for double copies between different types of
  special particles.  No explicit normalization is used.}

\defn{Mamp}{\vardefH{n}{Integer}}{Constructs the gravity tree amplitude
  using $1,\dots, n$ as the labels.}
~\\
\noindent It is important to note that \emph{only} \texttt{Aamp} includes
explicit normalizations.  However, we make the YMS normalization
accessible via

\defn{ymsNorm}{\vardefHH{traces}{List}}{Computes the YMS normalization
  $\mathscr{N}$ from \cref{eq:expl-norm}, with the trace groupings
  specified as seperate \texttt{Lists} in \varN{traces}.
  Gluons/fermions are specified via length-one
  \texttt{Lists}. \texttt{ymCoup} and \texttt{scCoup} are used to
  represent the sYM coupling $g$ and scalar coupling $\lambda$,
  respectively. }

\bibliographystyle{JHEP}
\bibliography{increasing_trees_bib}

\end{document}